\documentclass{aa}  
\usepackage{graphicx}
\usepackage{txfonts}
\usepackage{longtable}
\usepackage{lscape}
\usepackage{natbib}
\bibpunct{(}{)}{;}{a}{}{,}
\begin{document}
   \title{Multi-fibre optical spectroscopy of low-mass stars and brown dwarfs 
in Upper Sco \thanks{Based on observations obtained with the AAOmega spectrograph at the Anglo-Australian Observatory} \thanks{Table B.1 and optical spectra are only available in electronic form at the CDS via anonymous ftp to cdsarc.u-strasbg.fr (130.79.128.5) or via http://cdsweb.u-strasbg.fr/cgi-bin/qcat?J/A+A/}}

   \subtitle{}

   \author{N. Lodieu \inst{1,2}
          \and
          P. D. Dobbie \inst{3}
          \and
          N. C. Hambly \inst{4}
          }

   \offprints{N. Lodieu}

   \institute{Instituto de Astrof\'isica de Canarias (IAC), Calle V\'ia L\'actea s/n, E-38200 La Laguna, Tenerife, Spain\\
         \email{nlodieu@iac.es}
         \and
         Departamento de Astrof\'isica, Universidad de La Laguna (ULL),
E-38205 La Laguna, Tenerife, Spain
         \and
         Australian Astronomical Observatory, PO Box 296, Epping, NSW, 1710,
Australia\\
         \email{pdd@aao.gov.au}
         \and
         Scottish Universities' Physics Alliance (SUPA),
Institute for Astronomy, School of Physics, University of Edinburgh,
Royal Observatory, Blackford Hill, Edinburgh EH9 3HJ, UK \\
         \email{nch@roe.ac.uk}
             }

   \date{\today{}; \today{}}

 
  \abstract
   {Knowledge of the mass function in open clusters constitutes one way 
   to critically examine the formation mechanisms proposed to explain the 
   existence of low-mass stars and brown dwarfs.}
   {The aim of the project is to determine as accurately as possible 
the shape of the mass function across the stellar/substellar boundary 
in the young (5 Myr) and nearby (d = 145 pc) Upper Sco association.}
   {We have obtained multi-fibre intermediate-resolution (R$\sim$1100) 
optical ($\sim$5750--8800\AA{}) spectroscopy of 94 photometric and proper
motion selected low-mass star and brown dwarf candidates in Upper Sco with
the AAOmega spectrograph on the Anglo-Australian Telescope.}
   {We have estimated the spectral types and measured the equivalent 
widths of youth (H$\alpha$) and gravity (Na {\small{I}} and K {\small{I}}) 
diagnostic features to confirm the spectroscopic membership of about 95\% of the 
photometric and proper motion candidates extracted from 6.5 square 
degrees surveyed in Upper Sco by the UKIRT Infrared Deep Sky Survey (UKIDSS) 
Galactic Clusters Survey (GCS). We also detect lithium in the spectra
with the highest signal-to-noise, consolidating our conclusions about 
their youth. Furthermore, we derive an estimate of the efficiency of the
photometric and proper motion selections used in our earlier studies
using spectroscopic data obtained for a large number of stars falling
into the instrument's field-of-view.
We have estimated the effective temperatures and masses for each new 
spectroscopic member using the latest evolutionary models available 
for low-mass stars and brown dwarfs. Combining the current optical 
spectroscopy presented here with near-infrared spectroscopy obtained
for the faintest photometric candidates, we confirm the shape and slope
of our earlier photometric mass function. The luminosity function
drawn from the spectroscopic sample of 113 USco members
peaks at around M6 and is flat at later spectral type. We may detect 
the presence of the M7/M8 gap in the luminosity function as a result of 
the dust properties in substellar atmospheres. The mass function may peak 
at 0.2 M$_{\odot}$ and is quite flat in the substellar regime. We observe
a possible excess of cool low-mass brown dwarfs compared to IC\,348 and the 
extrapolation of the field mass functions, supporting the original 
hypothesis that Upper Sco may possess an excess of brown dwarfs compared 
to other young regions.}
   {This result shows that the selection of photometric candidates 
 based on five band photometry available from the UKIDSS GCS and 
 complemented partially by proper motions can lead to a good representation 
of the spectroscopic mass function.}

   \keywords{technique: spectroscopic --- stars: low-mass, brown dwarfs;
stars: luminosity function, mass function ---
galaxy: open clusters and associations: individual (Upper Scorpius)}

   \maketitle
%

%
%
\section{Introduction}

Firm knowledge of the form of the Initial Mass Function i.e.\ the number 
of objects as a function of mass \citep[IMF;][]{salpeter55,miller79, scalo86}
is vital to understanding the formation of low-mass stars and
brown dwarfs. To address fundamental issues like the universality
of the IMF (or otherwise its dependence with environment, time and/or
any other factor),
masses and effective temperatures (T$_{\rm eff}$) must be estimated
as accurately as possible for all members in open clusters and
star-forming regions. The advent of large optical Charge-Coupled Devices
(CCDs) and infrared detectors has facilitated the study of numerous
regions with different ages, densities, and histories.
However, the samples of photometric candidates confirmed 
spectroscopically in those regions for an unbiased estimate of the
IMF over a large mass range remains limited although these are growing rapidly.
Complete spectroscopic mass functions over the full mass range probed by the 
associated photometric surveys have been presented in the central region
of the Trapezium Cluster by \citet{luhman00b} and \citet{slesnick04}
and compared to photometric estimates \citep{hillenbrand00,muench02}.
Similarly, the Taurus region and the IC\,348 cluster have been targeted
photometrically and spectroscopically down to the substellar regime
\citep{luhman00a,briceno02,luhman03a,luhman03b}.
More recently, \citet{luhman07d} updated the census of low-mass
stars and brown dwarfs in Chamaeleon I star-forming region
based on optical and near-infrared spectroscopy.
The derived mass function in those regions would suggest a dependence
on environment of the IMF since the peak in the luminosity function 
of Taurus is at higher masses than in the Trapezium Cluster and IC\,348
\citep[0.8 vs 0.1--0.2 M$_{\odot}$;][]{luhman04a}.

The Upper Sco association (hereafter USco) is part of the Scorpius 
Centaurus OB association: it is located at 145 pc from the Sun 
\citep{deBruijne97} and its age is estimated to be 5$\pm$2 Myr from isochrone 
fitting and dynamical studies \citep{preibisch02}. The association was 
targeted in X-rays \citep{walter94,kunkel99,preibisch98}, with Hipparcos 
\citep{deBruijne97,deZeeuw99}, and more recently at optical 
\citep{preibisch01,preibisch02,ardila00,martin04,slesnick06} 
and near-infrared \citep{lodieu06,lodieu07a} wavelengths. 
As a result of the latest surveys, several tens of brown dwarfs were 
confirmed spectroscopically as members of the association
\citep{martin04,slesnick06,lodieu06,slesnick08,lodieu08a,martin10a},
resulting in a total of 92 substellar members divided into 76 M6--M9, 
and 16 L0--L2 dwarfs. \citet{slesnick08} presented the first spectroscopic
mass function in USco over the full area of the association down to 
M=0.02 M$_{\odot}$.
      
In this paper we present follow--up multi-fibre optical spectroscopy 
of low-mass star and brown dwarf candidates extracted from 6.5 square
degrees targeted by the UKIRT Infrared Deep Sky Survey
\citep[UKIDSS;][]{lawrence07} Galactic Clusters Survey (GCS) and 
published in \citet{lodieu07a}.
In Sect.\ \ref{USco_AAOmega:sample} we define the various samples
of targets included in our spectroscopic follow-up. 
In Sect.\ \ref{USco_AAOmega:spectro}, we describe the 
spectroscopic observations made with the AAOmega instrument installed 
on the 3.9-m Australian Astronomical telescope (AAT) and the data reduction.
In Sect.\ \ref{USco_AAOmega:Memb} we assess the membership of the
photometric candidates based on their spectral type, their chromospheric
activity, and the strength of gravity-sensitive features. A similar 
analysis is presented for a sample of photometric non-members to 
estimate the reliability of our original photometric and proper
motion selections in Sect.\ \ref{USco_AAOmega:Field}.
In Sect.\ \ref{USco_AAOmega:IMF} we derive effective temperatures,
bolometric luminosities, and masses for all USco members and compare
the spectroscopic luminosity and mass functions in the low-mass and 
brown dwarf regimes to previous studies in USco and other regions.
  
%
%
\section{The samples}
\label{USco_AAOmega:sample}

Our sample consists of three groups of objects: firstly photometric 
USco candidates selected from the UKIDSS GCS Science Verification
\citep{lodieu07a}; secondly, a few known spectroscopic members
falling in the instrument field-of-view and published in the literature 
\citep{martin04,slesnick06}, and, finally a sample of photometric 
non-members. 

The first sample consists of USco photometric candidates selected
from \citet{lodieu07a}. Out of the original 129 candidates, 114
are brighter than $Z$ = 17.5 mag, corresponding to masses in the
range 0.6--0.03 M$_{\odot}$ at the age and distance of USco. We were able 
to obtain spectra for 94 candidates out of 114, implying a spectroscopic 
completeness of 82\%. Table \ref{tab_USco:New_memb} lists the main 
properties of all 90 candidates targeted with AAT/AAOmega and subsequently 
confirmed as spectroscopic members, including coordinates,
$J$-band magnitudes\footnote{the full $ZYJHK$ photometry is published 
in \citet{lodieu07a}}, equivalent widths for the H$\alpha$ line and the 
Na {\small{I}} and K {\small{I}} doublets, spectral indices,
spectral types, effective temperatures, luminosities, and masses.
For the Na {\small{I}} and K {\small{I}} doublets we give the 
equivalent width of the doublet and not of each line that constitutes
the doublet although they are resolved at the resolution of our spectra.
For completeness, we have added to Table \ref{tab_USco:New_memb} the 
19 sources confirmed as spectroscopic members in the near-infrared 
with Gemini/GNIRS \citep{lodieu08a} and identified photometrically by 
\citet{lodieu07a}\footnote{23 candidates were presented in
\citet{lodieu08a}: 20 are from \citet{lodieu07a} and three from
\citet{lodieu06}. Two were classified as non members among the 20 candidates
from \citet{lodieu07a} but one is now reclassified as a member after optical
spectroscopy (see Sect.\ \ref{USco_AAOmega:Memb_contamination})}.
We should mention that five photometric candidates included in the
AAOmega survey are common to our earlier near-infrared spectroscopic 
follow-up \citep{lodieu08a}.

\begin{figure}
   \includegraphics[angle=0, width=\linewidth]{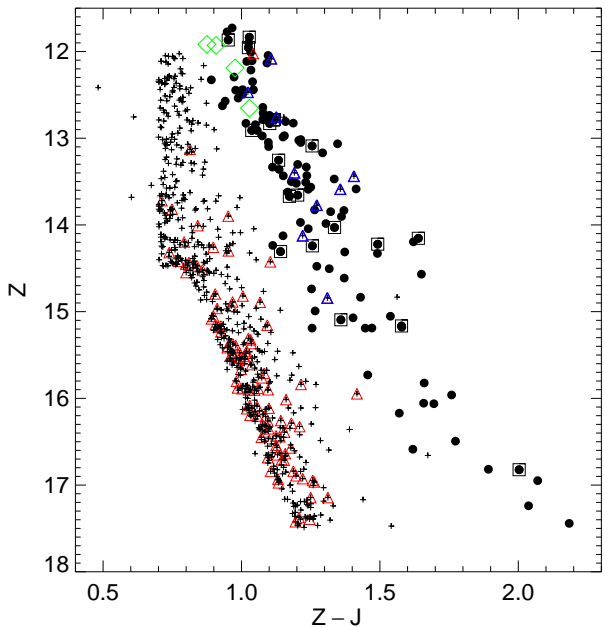}
   \caption{($Z-J$,$Z$) colour-magnitude diagram for all objects
in the AAOmega field-of-view (crosses) along with the new USco candidate 
members presented in this paper (filled dots) and photometric candidates 
lacking spectroscopy (open squares around filled dots). Overplotted are the 
nine objects classified as proper motion non-members by \citet{lodieu07a}
and reclassified as spectroscopic members here (blue open triangles).
Open red triangles represent a sample of stars in the AAOmega field
classified as photometric non members but with proper motions and equivalent 
width measurements consistent with young objects. Open green diamonds 
represent the four photometric candidates rejected as spectroscopic members 
after our AAOmega follow-up.
}
   \label{fig_USco:CMD_ZJZ}
\end{figure}

The second sample consists of five known spectroscopic members
with spectral types estimated from optical data with an accuracy
of half a subclass: DENIS1610$-$2212 \citep{martin04}, SCH1611$-$2217,
SCH1612$-$2349, SCH1612$-$2338, and SCH1613$-$2305 \citep{slesnick06}. 
These five objects fell in the AAOmega field-of-view and serve as 
spectral templates for assigning spectral types to our candidates. 
Table \ref{tab_USco:Known_memb} provides the full coordinates, $J$-band 
photometry, equivalent width measurements (in \AA{}) of the H$\alpha$ 
emission line and the gravity sensitive doublets (Na {\small{I}} and 
K {\small{I}}), as well as spectral types and names as quoted in 
the literature. The previously known members are not plotted in 
Fig.\ \ref{fig_USco:CMD_ZJZ} because no $Z$-band magnitudes are 
available for them as they lie outside the UKIDSS GCS field-of-view
\citep[see Fig.\ 4 in][]{lodieu07a}.

The third sample contains 624 sources we originally rejected as
members of USco based on their location in
various colour-magnitude diagrams \citep[Fig.\ 5 of][]{lodieu07a}.
Two subsamples were selected to match the magnitude range of the
USco photometric candidates. First, a bright sample
which consists of 269 sources with $Z$ = 12--14.5 mag and $Z-J$ 
colours redder than 0.7 mag (only 4 have $Z-J <$0.7 mag; 
Fig.\ \ref{fig_USco:CMD_ZJZ}). Second, a faint sample 
is made of 355 objects with $Z$ = 14.5--17.5 mag and selected to 
the right of a straight line going from ($Z-J$,$Z$) = (0.8,14.5) to
(1.2,17.5). These objects are shown as crosses in 
Fig.\ \ref{fig_USco:CMD_ZJZ}.

%
%
%
%

%
%
\begin{table*}
\caption{Known spectroscopic members in the AAOmega field used as templates.
References: (1) \citet{martin04}; (2) \citet{slesnick06}}
\label{tab_USco:Known_memb}
\centering
\begin{tabular}{@{\hspace{0mm}}c @{\hspace{0.2mm}}c c| c c c| c c c| c c c| c c| l@{\hspace{0mm}}}
\hline
\hline
RA & dec & $J$ & \multicolumn{3}{c|}{EW(H$_{\alpha}$)} & \multicolumn{3}{c|}{EW(NaI)} & \multicolumn{3}{c|}{EW(KI)} & SpT & new SpT & Name \\
h m s & $^\circ$ ' '' & mag & \multicolumn{3}{c|}{\AA{}} & \multicolumn{3}{c|}{\AA{}} & \multicolumn{3}{c|}{\AA{}} &  &  &  \\
\hline
16:10:50.00 & $-$22:12:51.8 & 12.680  & $-$15.9 & $-$10.9 & $-$7.0  & 4.1 & 3.3 & 3.2 & 4.2 & 3.2 & 2.8 & M5.5 & M5.5  & DENIS1610$-$2212$^{1}$ \\
16:11:17.11 & $-$22:17:17.5 & 14.340  & $-$11.0 & $-$3.6  &   ---   & 2.7 & 3.4 & --- & N/A & N/A & --- & M7.5 & M6.5  & SCH1611$-$2217$^{2}$ \\
16:12:37.58 & $-$23:49:23.4 & 13.930  & $-$7.9  & $-$7.5  & $-$9.4  & 5.0 & 4.6 & 4.7 & 3.8 & 3.7 & 3.6 & M6.0 & M5.5  & SCH1612$-$2349$^{2}$ \\
16:12:46.92 & $-$23:38:40.9 & 13.650  & $-$15.2 & $-$13.6 & $-$8.4  & 3.9 & 4.3 & 4.0 & 5.4 & N/A & 3.4 & M6.0 & M6.25 & SCH1612$-$2338$^{2}$ \\
16:13:12.12 & $-$23:05:03.3 & 14.050  & $-$5.7  & $-$11.6 & $-$1.6  & 3.6 & 4.4 & 3.6 & 3.9 & 5.3 & 4.8 & M6.5 & M6.5  & SCH1613$-$2305$^{2}$ \\
\hline
\end{tabular}
\end{table*}

%
%
\section{Spectroscopic follow-up of USco candidates}
\label{USco_AAOmega:spectro}
\subsection{Spectroscopic observations}
\label{USco_AAOmega:spec_obs}

We have obtained low-resolution multi-fibre optical spectroscopy of 94 
photometric and proper motion selected low-mass star and brown dwarf 
candidates in USco as well as a sample of photometric non members 
(Sect.\ \ref{USco_AAOmega:sample}) with the AAT/AAOmega spectrograph 
\citep{lewis02,sharp06}\footnote{More details on the AAOmega spectrograph 
can be found at the URL: http://www.aao.gov.au/AAO/astro/2df.html}. 
AAOmega has a two degree field-of-view which is sampled by 400 fibres that 
feed a two-armed spectrograph. The projected diameter on the sky of each 
fibre is 2 arcsec and their minimum separation is 30 arcsec 
\citep{miszalski06}. On the red arm we have employed the 385R grating to 
cover the wavelength range 5700--8800\,\AA{} (the exact coverage depends 
slightly on the position of the fibre on the spectrograph slit) at a 
resolution of R$\sim$1350\@. Spectra covering the wavelength range 
3740--5720\,\AA{}, also at a resolution of R$\sim$1350, were 
simultaneously obtained on the blue arm where we employed the 580V grating.

The observations were conducted during the nights 22--24 May 2007 by one of
us (PDD). Seeing and transparency were decidedly variable throughout this
period. The first night was dogged by substantial amounts of cloud and
poor seeing (2.5--3.0 arcsec). The next night was largely clear but seeing 
was again poor (2.5--3.0 arcsec). On the final night the sky was clear and 
the seeing was substantially better (0.8--1.5 arcsec). As this project was 
awarded an allocation of bright time, the moon was above the horizon for 
much of this period.

The observations were made with four fibre configurations at two
positions on the sky: the first field was centered approximately at
(RA,dec) $\sim$ (16$^{\rm h}$10$^{\rm m}$,$-$23$^{\circ}$)
whereas the second was positioned at
(RA,dec) $\sim$ (16$^{\rm h}$14$^{\rm m}$,$-$23$^{\circ}$10$'$).
Each field had a set-up for a sample of bright ($Z$ = 11.5--14.5 mag)
and faint ($Z$ = 14.5--17.5 mag) targets. Each field included science 
targets i.e.\ USco photometric candidates, USco known members confirmed 
spectroscopically by earlier studies, a sample of photometric non-members 
(or stars in the AAOmega field) picked-up randomly in the ($Z-J$,$Z$) 
colour-magnitude diagram as well as ``sky'' fibres (typically 50 per 
field). The UKIDSS images were checked to confirm that no sources 
was detected at the sky positions down to the detection limit of the GCS 
($J \sim$ 19.6 mag).

Single on-source integrations were set to 900 sec and 1800 sec for the
bright and faint samples, respectively. Total on-source integrations 
typically range from 75 min to 345 min for the bright sample and from 
270 min to 720 min for the faint sample. The total exposure time varies 
from target to target as $\sim$20 objects have only one spectrum and the 
others have two or more spectra (up to four). 
The data were reduced in the standard manner using the 2DFDR package
provided by the Australian Astronomical Observatory. We have separated the
various files (bias, flats, and objects) into directories and
performed an automated extraction of the one-dimensional spectra.
More details on the workings of the 2DFDR package are discussed in
\citet{reid10a}. We have measured 
the pseudo equivalent widths of the H$\alpha$ emission line and the 
gravity-sensitive features in each spectrum to look for variability. 
In summary, we present the combined spectra for 94 USco photometric 
and proper motion candidates, five previously known members, 
624 photometric non-members, and allocated 225 fibres to a sky position. 

%
%
\begin{table*}
\caption{Summary details of spectroscopic non-members, including 
coordinates, photometry, proper motions, and approximate spectral types}
\label{tab_USco:USco_NM}
\centering
\begin{tabular}{c c c c c c c c c c c}
\hline
\hline
RA & dec & $Z$ & $Y$ & $J$ & $H$ & $K$ & $\mu_\alpha \cos\delta$ & $\mu\delta$ & Type & Name \\
h s m & $^\circ$ ' '' & mag & mag & mag & mag & mag & mas/yr & mas/yr &  &  \\
\hline
16:13:15.65 & $-$23:27:44.2 & 11.918 & 11.548 & 11.043 & 10.279 & 10.053 &  $-$7.3 &  $-$8.7 & dK & cand8 \\
16:12:09.48 & $-$22:39:57.1 & 12.189 & 11.764 & 11.212 & 10.425 & 10.149 &   10.1  &  $-$7.8 & dK & cand13 \\
\hline
16:13:20.53 & $-$22:29:16.0 & 11.929 & 11.531 & 11.021 & 10.578 & 10.162 & $-$17.2 & $-$11.1 & M2 & cand7 \\
16:07:08.81 & $-$23:39:59.9 & 12.653 & 12.127 & 11.624 & 11.044 & 10.714 & $-$4.7  & $-$14.5 & M2 & cand23 \\
\hline
\end{tabular}
\end{table*}
\subsection{Spectral variability}
\label{USco_AAOmega:spec_VAR}

We observed some level of variability between spectra of the same source. 
Thirteen objects out 94 candidates identified in the GCS show one or 
two discrepant spectra. Similarly, we detect variability in SCH162$-$2338
among the five previously known members.
These variations may be caused by several effects: inhomogeneity
in the fibre response, chromatic variation in optical distortion, or 
intrinsic variability in the atmospheric properties of low-mass stars and 
brown dwarfs. We have obtained a total of 233 spectra for the 94$+$5=99 
sources and found that 13$+$1=14 objects are affected by this 
spectral variability, implying a rate below 15\%. Notably, most of the spectra 
affected show a significant level of fringing beyond 7500\,\AA{}.
The final spectral classification should not be 
affected by more than half a subclass because each object has generally 
two identical spectra and one or two discrepant. The discrepant datasets 
have been removed for the subsequent analysis.

%
%
%
\begin{figure}
   \includegraphics[angle=0, width=\linewidth]{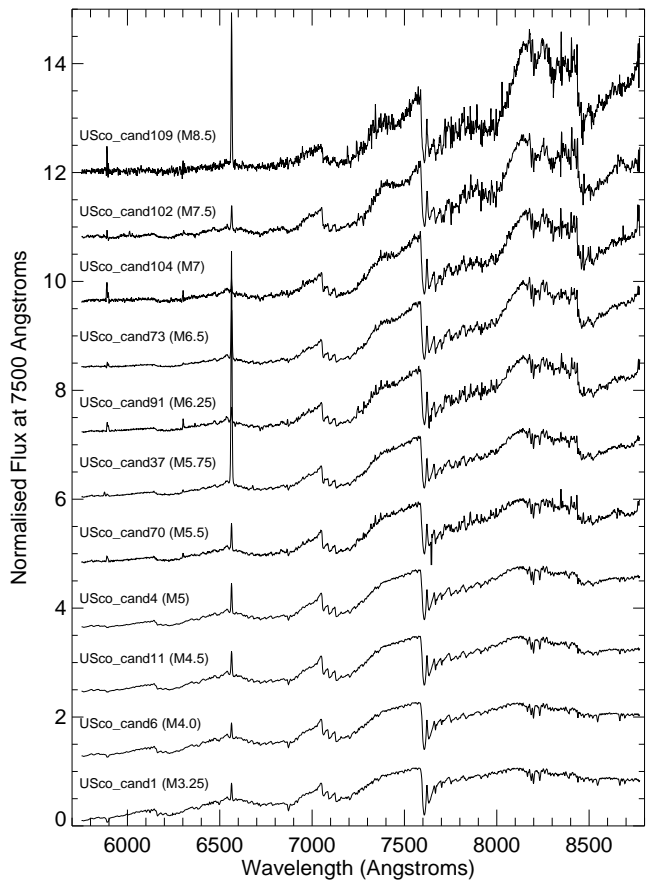}
   \caption{Combined optical spectra of USco members obtained with AAT/AAOmega.
The full range of spectral types (M3.25--M8.5) probed by our spectroscopic 
follow-up is shown with increasing spectral types from top to bottom. 
Spectra are shifted along the y-axis by increments of 1.2 for clarity.
}
   \label{fig_USco:all_spec}
\end{figure}
%
  
%
%
\section{Membership assessment}
\label{USco_AAOmega:Memb}
\subsection{Spectral types}
\label{USco_AAOmega:Memb_SpT}

Among our targets, we have included five previously known spectroscopically
confirmed members with spectral types ranging from M5.5 to M7.5
\citep[Table \ref{tab_USco:Known_memb};][]{martin04,slesnick06}.
These sources represent our ``primary'' templates as they were observed
with the same instrumental set-up as our photometric
candidates. The direct comparison of the spectra of these five members 
suggests that SCH1611$-$2217, SCH1612$-$2338, and SCH1613$-$2305 have 
similar spectral types and are cooler than DENIS1610$-$2212 and 
SCH1612$-$2349 whose spectra look very much alike 
(Table \ref{tab_USco:Known_memb}). 

In order to check their spectral types using the AAOmega 
spectra and to put our classification on an absolute scale, we have 
compared these five sources to young low-mass stars and brown dwarfs 
identified in the Chamaeleon I \citep[M3--M9;][]{luhman04b,luhman07d},
{$\eta$} Chamaeleon \citep[M3--M5.75;][]{luhman04f} and 
Taurus \citep[M5.25--M9.5;][]{briceno98,luhman03a} star-forming regions,
and the Oph1622$-$2405 binary \citep[M7.25 \& M8.75][]{luhman07b}.
The grid of spectral types available is fairly complete from M3 to M9.5
with a template every 0.25 subclass as follows: M3, M3.25, M4, M4.5,
M4.75, M5, M5.25, M5.5, M5.75, M6, M6.25, M6.5, M7.25, M7.5, M8.25, M8.5,
M8.75, and M9.5\footnote{Spectra kindly provided by Dr.\ Kevin Luhman}.
We have revised the spectral types of the five USco members as indicated in 
Table \ref{tab_USco:Known_memb} by comparing them directly to the templates 
aforementioned. Uncertainties on the spectral types are typically given by 
the spacing of the grid i.e.\ 0.25--0.5 subclass in most cases. The new 
spectral types match the previous estimates for known USco members within 
the uncertainties quoted by the earlier studies, 
except for SCH1611$-$2217 for which we derived M6.5 instead of M7.5 
\citep{slesnick06}. We favoured this direct comparison over the
determination of spectral types using spectral indices defined in
the literature for field stars \citep[e.g.\ the PC3 index;][]{martin96}
because they may not be reliable for young low-mass stars and brown dwarfs.

Among our sample of 94 GCS candidates, we found two objects
of spectral types earlier than M, arguing that they are spectroscopic non 
members, UGCS J161315.65$-$232744.2 and UGCS J161209.48$-$223957.1 
(Table \ref{tab_USco:USco_NM}). To assign spectral types to the remaining 
92 candidates, we compared their combined spectra to our spectral templates: 
SCH1612$-$2305 \citep[M6.5;][]{slesnick06} and DENIS1610$-$2212 
\citep[M5.5;][]{martin04}. Then, we have compared the optical spectra to
each other to create groups following a spectral type sequence. We have assigned
a spectral type per group using young templates in Chamaeleon, Taurus, and 
Ophiuchus. We have found two sources
with M2 spectral type which appear too faint for their spectral types. For 
this reason we consider them as non members in the rest of the paper.
The spectral types span the M3.25--M8.5 range and are listed in 
Table \ref{tab_USco:New_memb}. 

Columns 1 and 2 of Table \ref{tab_USco:New_memb} provides the coordinates 
(J2000) and $J$-band magnitudes of 90 new spectroscopic candidate members of 
the USco association. Columns 3--6, 7--10, and 11--14 give the measurements 
(up to four) of the pseudo-equivalent widths (in \AA{}) of the H$\alpha$ 
line as well as the Na {\small{I}} and K {\small{I}} doublets, respectively,
discussed in the following sections. We emphasise here that membership
of the USco association is assigned based on the various criteria
available to us: photometry, proper motion, H$\alpha$ emission, 
gravity-sensitive features, lithium and magnesium lines when ``available''.
The presence of contaminants cannot be completely ruled out but their level 
should be extremely low when using a combination of all these criteria.
A long dash line indicates that either no spectrum is available or
the signal-to-noise ratio is too low to publish a reliable measurement.
Columns 15--18 list spectral indices; Columns 19$+$20 provide spectral
types derived from the spectral indices and estimated from the spectral
templates, respectively. Spectral types are accurate to 0.25 class for
those earlier than M7.5 and half a subclass for later objects.
Columns 21--23 provide the estimated effective temperatures (in K),
luminosities (log(L$_{\rm bol}$/L$_{\odot}$)), and masses (in M$_{\odot}$).

%
%
%
\begin{figure}
   \includegraphics[angle=0, width=\linewidth]{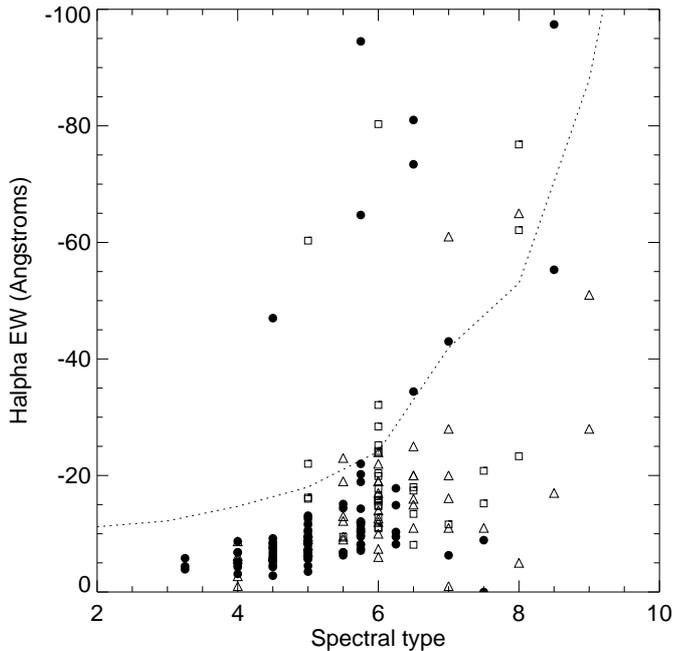}
   \caption{H$\alpha$ equivalent widths (EWs; in \AA{}) as a function of
spectral type (4$\equiv$M4; 5$\equiv$M5; etc\ldots{}) for all photometric
candidates confirmed as spectroscopic members (filled dots) in this study.
Three objects lie outside this plot due to their extremely strong
H$\alpha$ emission. We have added data points from earlier
studies for comparison: open triangles come from \citet{ardila00} and
\citet{martin04} and open squares are from \citet{slesnick06}.
}
   \label{fig_USco:Ha_SpType}
\end{figure}
%

%
%
%
\begin{figure*}
   \centering
   \includegraphics[angle=0, width=0.49\linewidth]{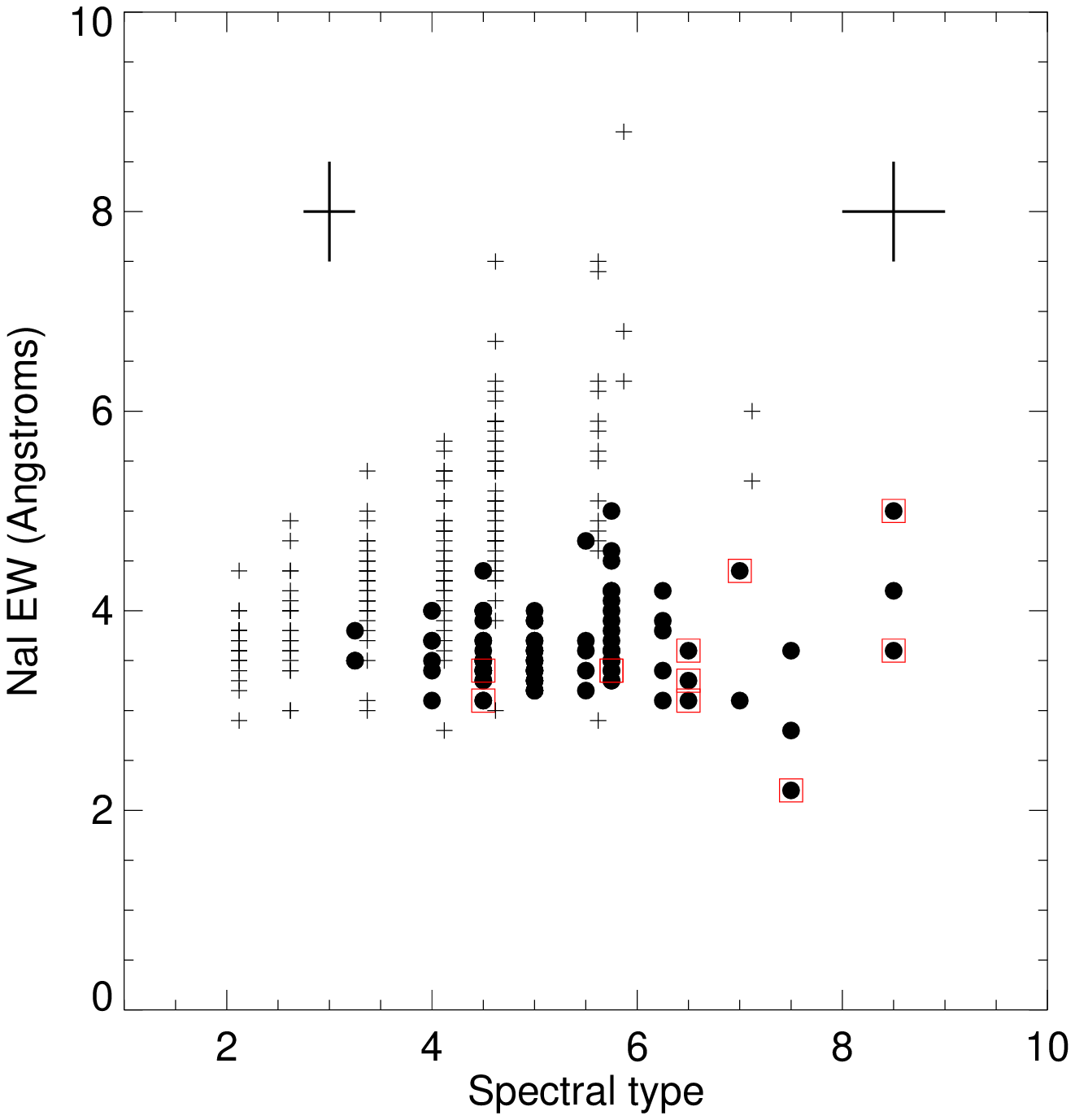}
   \includegraphics[angle=0, width=0.49\linewidth]{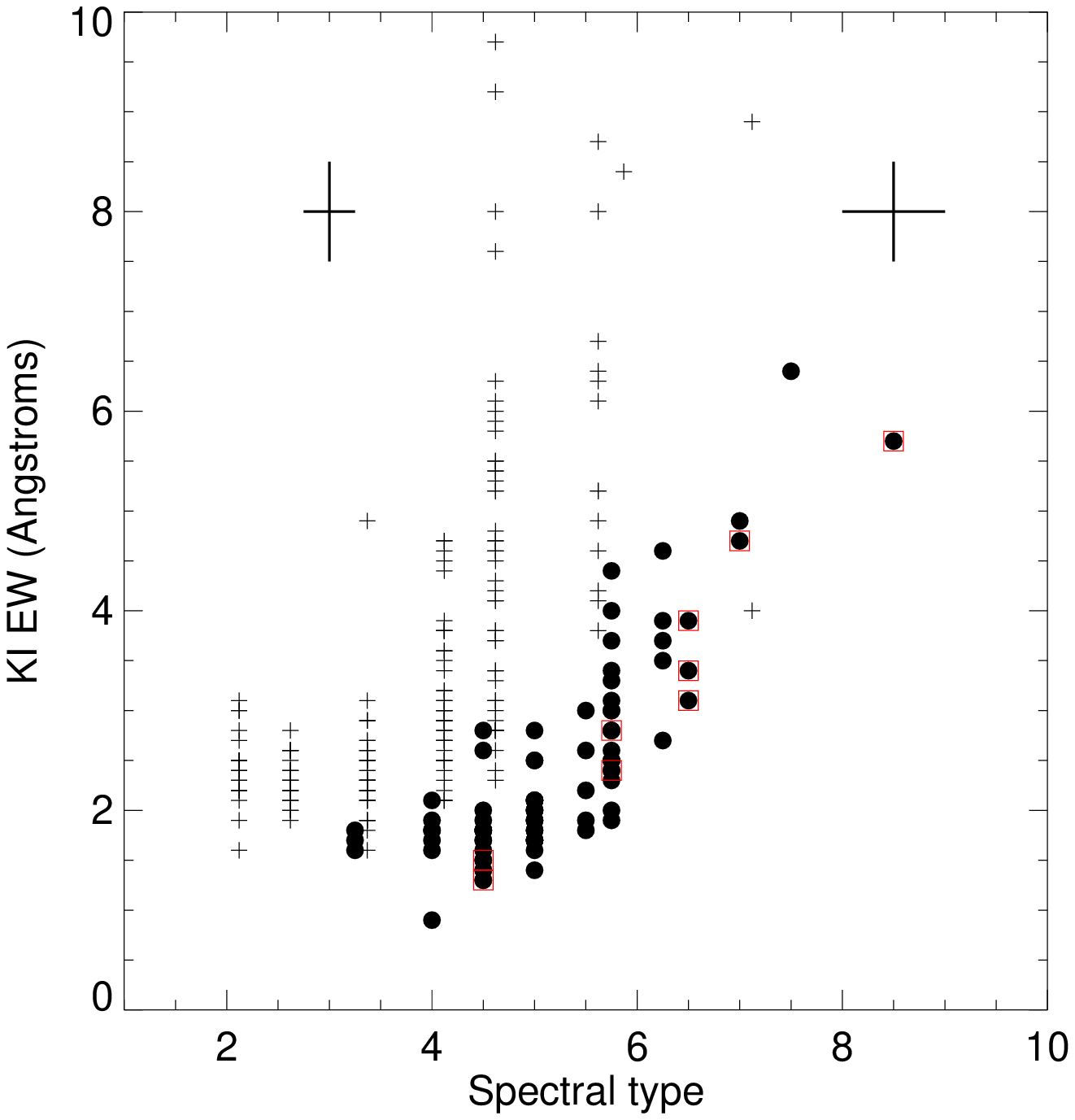}
   \caption{Na {\small{I}} (left) and K {\small{I}} (right) equivalent width
measurements (in \AA{}) as a function of spectral type (4$\equiv$M4;
5$\equiv$M5; etc\ldots{}) for all the photometric candidates confirmed as
spectroscopic members. {{Typical error bars on the equivalent width
measurements are marked.}} Some of the latest type members lack measurements
for the K{\small{I}} doublet because of the nearby strong oxygen telluric
band. The candidates for accretion i.e.\ which exhibit H$\alpha$
equivalent widths larger than the empirical boundary defined by
\citet*{barrado03b} are highlighted with red squares. The stars lying
in the AAOmega field have been added as crosses and shifted in
spectral types by $+$0.12 for clarity.
}
   \label{fig_USco:NaI_KI_SpType}
\end{figure*}
%

%
%
%
\begin{figure*}
   \centering
   \includegraphics[angle=0, width=0.49\linewidth]{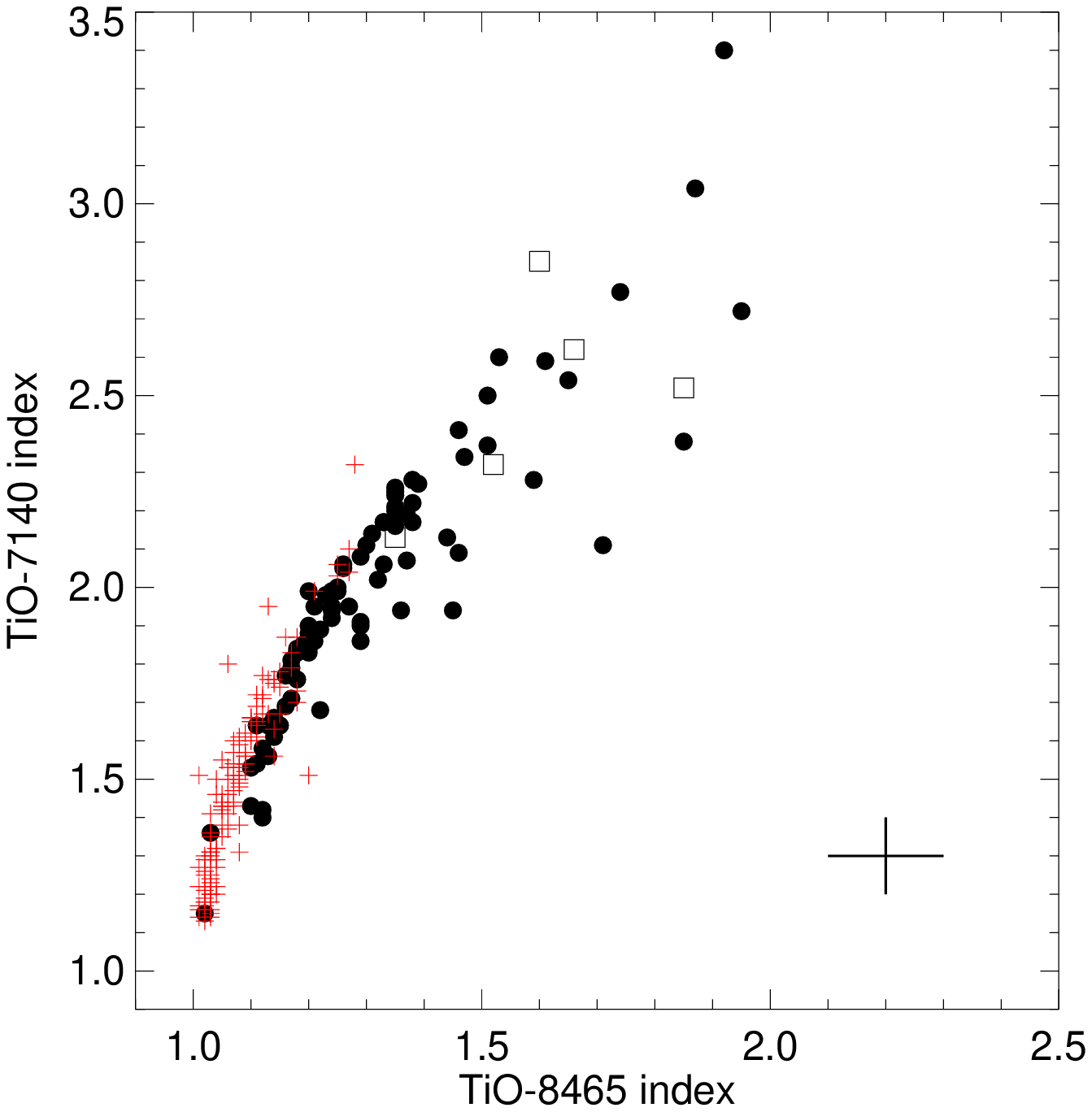}
   \includegraphics[angle=0, width=0.49\linewidth]{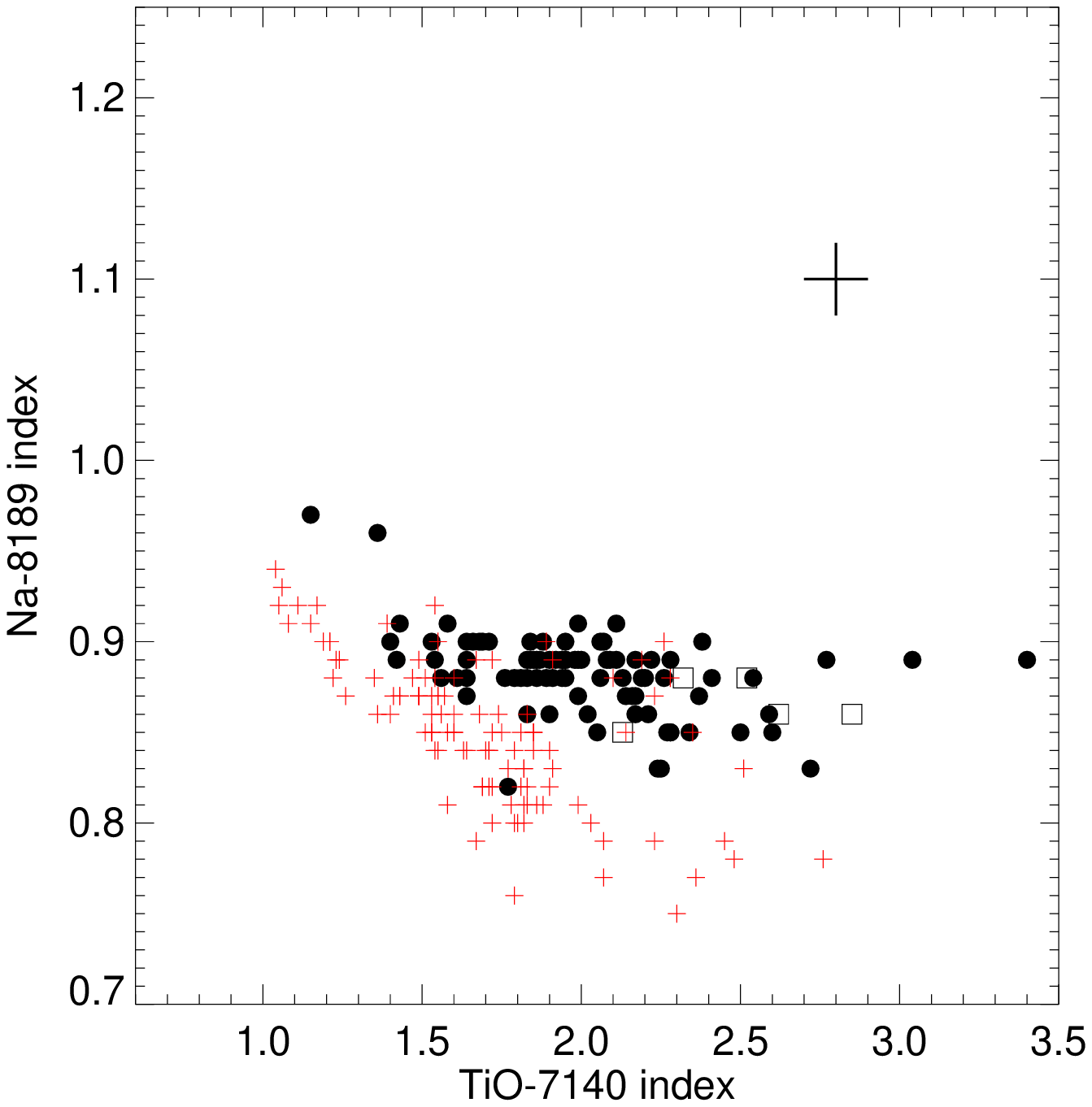}
   \caption{Temperature (left) and gravity (right) sensitive indices
for the new USco candidate members (filled circles), previously known
members (open squares), and stars in the AAOmega field-of-view (crosses).
Typical error bars are shown.
Plots originally proposed by \citet{slesnick06}.
}
   \label{fig_USco:indices}
\end{figure*}
\subsection{Chromospheric activity}
\label{USco_AAOmega:Memb_Halpha}

In this section, we measure the equivalent widths of the H$\alpha$ line
at 6563\,\AA{} for the 90 candidates whose optical spectra substantiate our 
claims about their membership of the USco association. The detection of strong 
H$\alpha$ in 
emission in the optical spectral of young low-mass stars and brown dwarfs 
is generally assumed to indicate youth. We used the SPLOT task under
IRAF\footnote{IRAF is distributed by National Optical Astronomy Observatory, 
which is operated by the Association of Universities for Research in 
Astronomy, Inc., under contract with the National Science Foundation.}
to measure pseudo-equivalent widths to an accuracy of 1\,\AA{} or better
(depending on the strength of the line). M dwarfs are long-lived and
remain active for a long time. A large fraction of M3--M9 field dwarfs
are active, on average 40\% but with a range from 20\% for M3
up to $\sim$100\% for the latest M dwarfs \citep{west08}. However, only
$\sim$15\% of old field M dwarfs exhibit strong H$\alpha$  
emission with equivalent widths less than $-$5\,\AA{}.

We detect H$\alpha$ in emission in the optical spectra of the USco
candidate members (Fig.\ \ref{fig_USco:Ha_SpType}), except one source for 
which H$\alpha$ is in absorption despite its late spectral type, 
UGCS J161340.79$-$221946.1 (M7.5)\@. Absorption lines occur for
stars in standard evolutionary states such main-sequence or post-main
sequence but are unlikely in the pre-main sequence stage. However, due
to the large amount of fringing and the low signal-to-noise of this 
particular spectrum, we prefer to keep it as a possible candidate.
We also observe different levels 
of variability in our equivalent width measurements reported in
Table \ref{tab_USco:New_memb}. As an example, we have three sources,
UGCS J161216.09$-$234425.0, UGCS J161126.30$-$234006.1,
and UGCS J161047.13$-$223949.4, where we detect H$\alpha$ in emission
in only two of the four spectra. In particular, the latter shows strong
H$\alpha$ emission with equivalent widths of $\sim$$-$55 and $\sim$$-$30\,\AA{}
while no emission is detected in the other two spectra. Nonetheless, 
we keep these sources as members because their spectra are otherwise consistent
with young low-mass stars. These measurements show that variability
is common among young low-mass stars and brown dwarfs and that the 
absence of H$\alpha$ alone is not sufficient to rule out youth
as a characteristic of low-mass stellar and substellar objects
\citep{martin10a}.
 
The H$\alpha$ emission can be due to, on the one hand, chromospheric 
activity, and, on the other hand, to accretion or strong winds.
The latter is usually inferred from the strong levels of H$\alpha$
emission. To quantify this statement, \citet*{barrado03b} have defined
an empirical boundary between non-accreting and accreting M-type stars and
brown dwarfs. We have identified 11 new accreting candidates that lie 
significantly above this empirical boundary (dotted line in
Fig.\ \ref{fig_USco:Ha_SpType}). One source, UGCS J160648.18$-$223040.1, 
lies outside the limits of the plot displayed in 
Fig.\ \ref{fig_USco:Ha_SpType} because of its extremely strong H$\alpha$ 
emission. In the first spectrum, we have measured an equivalent width
(EW) of $-$1780\,\AA{} whereas the emission is weaker (EW = $-$730\,\AA{})
in the second spectrum although as strong as the strongest H$\alpha$
emission detected in a member of the young $\sigma$Orionis cluster
\citep[SOri\,71; EW = $-$705\,\AA{}][]{barrado02b}. We have identified
another object, UGCS J160723.82$-$221102.0 which also exhibits strong
H$\alpha$ emission with equivalent width measurements of 
$-$671.0 and $-$431.5\,\AA{}, comparable to SOri\,71\@. Moreover,
we have measured H$\alpha$ equivalent widths of order $-$155 to $-$135\AA{} 
for USco J161354.34$-$232034.4\@.

\subsection{Surface gravity}
\label{USco_AAOmega:Memb_grav}

To further assess the membership of our USco photometric candidates,
we have measured the pseudo-equivalent widths of the
gravity-sensitive Na {\small{I}} (8183/8195\,\AA{}) and K {\small{I}}
(7665/7699\,\AA{}) doublets in each individual spectrum 
(Table \ref{tab_USco:New_memb}).

The left-hand side panel of Fig.\ \ref{fig_USco:NaI_KI_SpType} 
shows the behaviour of the equivalent widths of the Na {\small{I}} doublet 
as a function of spectral type. Its equivalent width appears constant as 
a function of spectral type with a mean value of 3.5--4\,\AA{} and a 
dispersion of 1\,\AA{}. The upper envelope for Upper Sco candidate members 
is around 5.5\,\AA{} as seen to Figure 4 in \citet{martin04}, consistent 
with the trend of our measurements (we find an upper limit of 5\,\AA{}). 
The trends in open clusters like the
Pleiades \citep{martin96} and Alpha Per \citep{lodieu05a} indicate mean 
values of 5$\pm$2\,\AA{} for the equivalent widths of the Na {\small{I}} 
doublet in the M4--M8 spectral type range whereas old field dwarfs have 
equivalent widths typically larger than 6\,\AA{} \citep{martin96}.
Therefore, our measurements are consistent with general trends and
consolidate our conclusion that our USco photometric candidates are 
young objects.

The right-hand side panel of Fig.\ \ref{fig_USco:NaI_KI_SpType}
shows the behaviour of the K {\small{I}} doublet as a function of
spectral type for the new USco candidate members. The equivalent
widths of the doublet exhibits an increase as a function of spectral 
type and values in the 2--6\,\AA{} range.

On both panels of Fig.\ \ref{fig_USco:NaI_KI_SpType}, we have overplotted 
the equivalent width measurements of 180 stars lying in the AAOmega 
field-of-view and classified as M dwarfs without H$\alpha$ in emission 
(red crosses). The mean values of the equivalent widths of the Na {\small{I}}
and K {\small{I}} doublets for these field M dwarfs is clearly higher than 
the average of the new USco candidate members for each spectral type. 
One hundred and twenty seven ($\sim$70\%) stars in the AAOmega field have 
Na {\small{I}} doublet equivalent widths larger than 4\,\AA{}, pointing 
towards an older age on average at a given spectral type than for our USco 
candidate members. The difference is even more marked for the K {\small{I}}
doublet.

We have also computed the temperature (TiO-$\lambda$7140 and
TiO-$\lambda$8465) and gravity (Na-8189) sensitive indices designed
by \citet{slesnick06} for the new USco candidates (filled circles) and 
previously known members (open squares in Fig.\ \ref{fig_USco:indices}). 
Figure \ref{fig_USco:indices} shows the same plots with the same (x,y)
scales as in Figure 2 of 
\citet{slesnick06} and Figures 10--11 of \citet{slesnick08} for direct 
comparison between their samples and ours. Note that the six faintest
sources of our sample of candidate members are not included in 
Fig.\ \ref{fig_USco:indices} because of strong fringing beyond
8000\,\AA{} affecting the computation of the TiO-$\lambda$8465
and Na-8189 indices. Our values of the three indices agree within 
10\% with their values published in their Table 1\@.
The trend in the left-hand side plot of Fig.\ \ref{fig_USco:indices} 
confirms the range in spectral types (M3.25--M8) derived from direct 
comparison with young templates. A few objects lie below the sequence, 
a position attributed to the occurence of veiling by \citet{slesnick06}. 
Similarly, the mean values of the Na-8189 for our sample is around
0.9 with a dispersion of 0.3, consistent with the range of values
($\sim$0.86--0.96) reported by \citet{slesnick06} and \citet{slesnick08}.

The positions of the stars in the AAOmega field-of-view in the 
left-hand side diagram of Fig.\ \ref{fig_USco:indices}
follow the general trend of the USco candidates mainly because this
diagram is independent of the age (i.e.\ gravity). However, the locations
of these objects in the right-hand side diagram differ from the positions
of the USco candidates and appear intermediate between field dwarfs
and pre-main-sequence stars (blue and green crosses in Figure 2 of
\citet{slesnick08}, respectively). This trend points towards an 
intermediate age for the stars in the AAOmega field and corroborates
the conclusions drawn from the equivalent widths of the Na {\small{I}} 
doublet. Their position in the ($Z-J$,$Z$) colour-magnitude diagram
displayed in Fig.\ \ref{fig_USco:CMD_ZJZ} indicates that these objects 
are, on average, fainter than the USco members for the same spectral type,
suggesting that they are located beyond the USco association.

\subsection{Other features: lithium and magnesium}
\label{USco_AAOmega:Memb_Li}

\begin{figure}
   \includegraphics[angle=0, width=\linewidth]{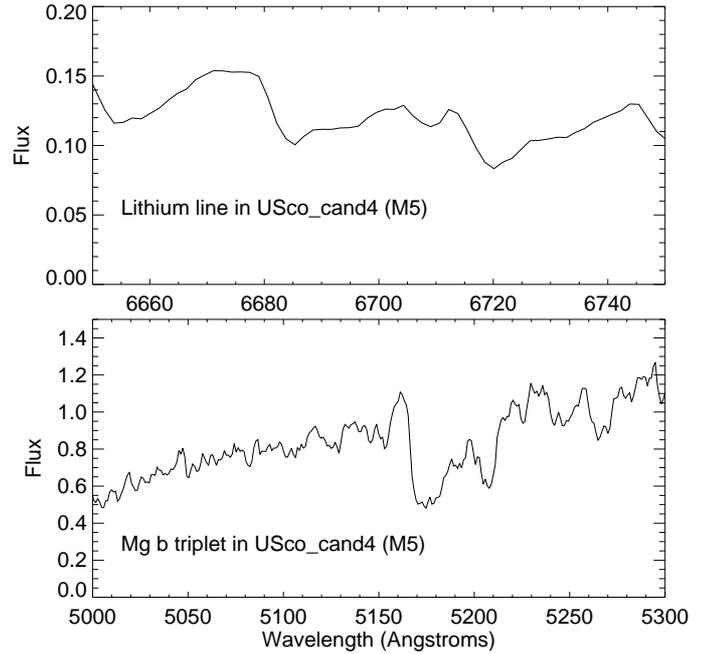}
   \caption{Region around the magnesium triplet at $\sim$5200 Angstroms
(bottom) and the lithium line at 6708.8\AA{} for one USco members
(USco\_cand4; M5).}
   \label{fig_USco:spec_Mg_Li}
\end{figure}

Although searching for the presence or absence of the lithium absorption 
line at 6707.8\,\AA{} was not originally an aim of this project, we are 
able to detect that line thanks to a combination of the moderate resolution 
(R$\sim$1100) and the high quality of the spectra obtained for the brightest 
sources ($J \leq$13 mag and $Z \leq$14.5 mag). The strength of the
equivalent width varies from object to object and is of the order of
0.2--0.5\,\AA{}. The lithium is detected in absorption in almost all 
photometric candidates confirmed spectroscopically on the basis of the 
presence of H$\alpha$ and weak gravity features (top panel in 
Fig.\ \ref{fig_USco:spec_Mg_Li}), again adding credence to their inferred 
youth and membership of the USco association. 

In the blue part of the spectra, we detect the presence of the Mg b
triplet around 5200\,\AA{} with an excellent signal-to-noise for the
brightest USco members (bottom panel in Fig.\ \ref{fig_USco:spec_Mg_Li}). 
The full spectral range provided by the blue grating (3740--5720\,\AA{}) 
at a resolution of R$\sim$1350 could be used to study the accretion rates 
in a large sample of young low-mass stars and brown dwarfs at 5 Myr 
\citep[e.g.][]{herczeg09} but such a study is beyond the scope of 
this paper.

%
%
%
\subsection{Level of contamination}
\label{USco_AAOmega:Memb_contamination}

Among the 94 photometric and proper motion candidates followed-up 
spectroscopically with AAOmega, we have confirmed the membership of 
90 translating into a success rate of the original selection of $\sim$96\%. 
Two contaminants listed in Table \ref{tab_USco:USco_NM} appear 
to be background stars and the other two field early-M dwarfs. 
This result demonstrates the performance of the photometric and 
astrometric selections described in \citet{lodieu06} and \citet{lodieu07a}.

We note that four objects were previously confirmed through their
near-infrared cross-dispersed spectra \citep{lodieu08a}. The
spectral types of USco J160830.49$-$233511.0 (M8.5 in the optical
vs M9 in the near-infrared), USco J160648.18$-$223040.1 (M8.5 vs M8),
and USco J161047.13$-$223949.4 (M8.5 vs M9) agree within a subclass.
The optical and near-infrared spectral types of USco J160723.82$-$221102.0 
differ by several subclasses (M7.5 vs L1) as noted by \citet{herczeg09}
who independently classified it as a M8.5 dwarf. Another object,
USco J161421.44$-$233914.8, rejected by our near-infrared spectroscopic
follow-up, is now reclassified as a member with a M7 spectral type
according to the AAOmega spectrum. We have checked the input coordinates and 
positions in both runs and cannot find out the reason for this discrepancy.
Unfortunately, we could not reach fainter candidates to confirm our
near-infrared classification with optical spectroscopy. The former
has been criticised by \citet{herczeg09} on the basis of only two sources 
where the classifications are discrepant. We should point out that younger 
M dwarfs are hotter than their field counterparts, implying that the vanadium 
oxide (VO) band should remain at later spectral types for younger ages and 
thus lead an earlier spectral classification \citep{martin96}.

For the faintest USco candidates with estimated masses below
0.03 M$_{\odot}$ according to theoretical models, our earlier study
showed that 19 out of 20 photometric candidates are indeed spectroscopic
members \citep{lodieu08a}. Their near-infrared cross-dispersed spectra
confirm them as ultracool dwarfs with features characteristics of young 
objects and effective temperatures in the 2700--1800\,K range.
Therefore, the level of contamination in the 0.03--0.006 M$_{\odot}$ mass 
range is of the order of 5\% after reclassifying 
USco J161421.44$-$233914.8. Over the full magnitude and mass ranges
probed in USco by the GCS, our success rate is $\sim$95\%.
  
%
%
\section{Probing the efficiency of the original selection}
\label{USco_AAOmega:Field}

In addition to the USco members discussed in 
Sect.\ \ref{USco_AAOmega:Memb_SpT}, we assigned 624 fibres to sources 
located to the blue side of the lines used for our photometric selection 
objects referred as to ``stars in the AAOmega field''). 
Among these extra sources, we have identified a large number of stars 
(Sect.\ \ref{USco_AAOmega:Field_sample}), proper motion non members, as 
well as three accreting sources (Sect.\ \ref{USco_AAOmega:Field_accreting}). 
We have used this large sample to estimate the efficiency of the original 
photometric (Sect.\ \ref{USco_AAOmega:phot_selection}) and proper motion 
(Sect.\ \ref{USco_AAOmega:PM_selection}) selections described in 
\citet{lodieu07a} based on multiple criteria, including the magnitudes, 
spectral types, strengths of the equivalent width of activity and 
gravity features.

\subsection{The sample of stars in the AAOmega field}
\label{USco_AAOmega:Field_sample}

We have assigned fibres to over 600 stars or point sources not 
selected by our photometric study as members of the USco association
\citep[crosses in Fig.\ \ref{fig_USco:CMD_ZJZ};][]{lodieu07a}. We 
provide the coordinates, photometry, astrometry and tentative spectral 
types for all stars in the AAOmega field (Table \ref{tab_USco:field_stars}) 
as well as the optical 
spectra to the community\footnote{Tables \& AAOmega optical spectra taken 
with the blue (3740--5720\,\AA{}) and red (5600--8800\,\AA{}) gratings are 
available at www.iac.es/galeria/nlodieu/publications/} as they may be 
useful to future surveys of the region or to other type of scientific 
exploitation. The large majority of these objects do not belong to the 
Upper Sco association, except the ones discussed in the following
sections and listed in Table \ref{tab_USco:PM_NM_L07a}.

We have assigned rough spectral types accurate to a few spectral classes
using a combination of criteria including direct comparison with
low-resolution spectra downloaded from the European Southern Observatory
database\footnote{www.eso.org/instruments/isaac/tools/lib/index.html}
and expected strengths of major lines and absorption bands 
(e.g.\ NaD, TiO, CaH, Ca {\small{I}}, Na {\small{I}}) as a function of 
the luminosity class and spectral types (Table 5 of \citet{kirkpatrick91} 
and spectra in \citet{turnshek85}).

The full sample of spectra can be divided up into several groups for
which we have chosen one template because of the high quality of its
spectrum. The majority of objects enters in the class of M dwarfs
with spectral types later than M2 (right panel of 
Fig.\ \ref{fig_USco:spec_others}). 
We have also identified a large number of late-K/early-M dwarfs 
(K7--M1) with different levels of reddening and late-G giants
(left-hand side of Fig.\ \ref{fig_USco:spec_others}). Finally,
we have classified three sources as young active flare M dwarfs
(details in Sect.\ \ref{USco_AAOmega:Field_accreting}), two white 
dwarfs, 18 objects that we were unable to classify due to problems 
with the fibre response or low signal-to-noise, and 11 sources have
no signal. Tables, plots and selected templates are detailed in 
Appendix \ref{USco:appendix_field_stars}.

\subsection{Contamination in the proper motion selection}
\label{USco_AAOmega:PM_selection}

In \citet{lodieu07a}, we used a 2$\sigma$ clip to reject proper motion 
non members, implying that our selection should have included $\sim$95\% 
of members but missed $\sim$5\% of true members (assuming normally 
distributed errors). In this section, we aim to address the question related 
to the level of contamination present in our original cut in 
the proper motion selection: do we find that 5\% of the proper motion non 
members listed in Table C1 of \citet{lodieu07a} are actually members? 

In \citet{lodieu07a}, we discarded 23 photometric candidates as proper
motion non members\footnote{3 sources in Table C1 of \citet{lodieu07a} are 
common to the sample of photometric non members listed in their Table B1}.
Ten of these 23 proper motion non members were included in our spectroscopic 
follow-up having been assigned a fibre (Table \ref{tab_USco:PM_NM_L07a}). 
We have looked at their spectral types, H$\alpha$ and Na {\small{I}} 
equivalent widths, and the presence of lithium to investigate 
their membership. Among these 10 sources, only one is rejected as
a member based on its high Na {\small{I}} equivalent width.
The remaining nine objects show H$\alpha$ in emission and weak
gravity-sensitive features, and have spectral types between M4 and M6,
consistent with young objects. Among those nine sources, six have a clear
lithium detection while the other three may exhibit lithium in absorption
but the detection is only tentative.

The results, detailed above, are summarised in Table \ref{tab_USco:PM_NM_L07a}.
Column 1 of Table \ref{tab_USco:PM_NM_L07a} gives the name of the 
object. Column 2 provides the $Z$-band magnitude; Columns 3 \& 4 the 
proper motion (mas/yr) in right ascension and declination, respectively;
Column 5 the spectral types; Columns 6 \& 7 H$\alpha$ and Na {\small{I}} 
equivalent widths (EWs in \AA{}), both individual measurements and the sum;
Column 8 the presence of lithium (Y$\equiv$Yes); Column 9 a note on 
the proper motion, Column 10 the final decision on the membership, and
the last column the proper motions from UCAC3 \citep{zacharias10a}.

We have checked the UCAC3 catalogue for the other 13 sources classified as 
proper motion non members in Table C1 of \citet{lodieu07a}. Seven objects 
brighter than $Z \leq$ 13 mag have a UCAC3 counterpart and astrometry accurate 
to a few mas/yr \citep{zacharias10a}. The quality of the astrometry provided 
by UCAC3 supersedes our previous proper motion measurements and is useful to 
constrain membership in USco \citep{bouy09c}. Three are clearly proper motion 
non members: USco J160937.85$-$212319.0, and the (possible) wide binary 
composed of USco J161002.67$-$234439.5 and USco J161002.86$-$234440.9\@.
Another object, USco J161626.20$-$235048.8, is possibly not a member.
The other three bright objects are proper motion members but lack
optical spectra because they were not assigned a fibre during our
spectroscopic follow-up. The remaining six (out of 13) are not in
the UCAC3 online catalog mainly because they are too faint.

We have made a comparison between the proper
motions in right ascension (pmRA) and declination (pmDEC) from UCAC3 
and the ones derived from the 2MASS/GCS cross-match \citep{lodieu07a}.
This sample of stars with $J$ = 10.75--13 mag contains 2815 sources.
The agreement between both measurements is usually good with Gaussian
equivalent root-mean-squares of 8.04 and 7.95 mas/yr in right ascension
and declination, respectively. For the 43 USco candidates with UCAC3
measurements, we find median absolute deviations of 11.1 and 6.53 mas/yr
in right ascension and declination, respectively. These error bars
estimated from both datasets appear very similar in size.

To summarise, we find that our 2$\sigma$ clip in the proper motion
mis-classified 12 members (out of 17) as proper motion non members 
(and possibly up to 18 out of 23). In Lodieu et al. (2007b) we
rejected 23 out of 139 photometric candidates as proper motion 
non members because they lay beyond the 2$\sigma$ cut off. For
normally distributed errors, we would expect 5\% of 139 = 7
photometric members to be outside 2$\sigma$ just by chance. Here
we find that between 12 and 18 objects are actually members
i.e.\ a factor of two larger than the nomical 5\%. Only one
object, USco J161412.41$-$221913.3, does lie outside the 3$\sigma$ 
circle centered on the cluster mean proper motion
(Table \ref{tab_USco:PM_NM_L07a}), suggesting that our original selection 
was pragmatic and threw away some bona-fide members. Future astrometric
selection should rather employ 3$\sigma$ clips to optimize the selection
of members. The uncertainties in the proper motion measurements
from the 2MASS/GCS cross-match was typically 10 mas/yr down to 
$J$ = 15.5 mag. We hope to improve the reliability of the astrometry 
by combining the UKIDSS GCS with the second epoch planned within the
framework of the VISTA public surveys \citep{emerson01}\footnote{More
details on VISTA at www.vista.ac.uk}.

\subsection{Completeness of the selection}
\label{USco_AAOmega:phot_selection}

In this section, we discuss the completeness of the original 
photometric selection designed to identify members in the USco association
\citep{lodieu07a}. To test this original photometric selection, we have 
computed lower and upper limits for the completeness by comparing 
the number of photometric candidates followed-up spectroscopically to
the 129 photometric candidates identified in the GCS after pure
photometric selection \citep{lodieu07a}. A total of 94 sources
have optical spectra and four are non members while 20 were observed
in the near-infrared (one is classified as a non member). The remaining
sources currently lack spectroscopic follow-up. Therefore,
the lower limit for the efficiency of our selection is $\sim$81\%
(=(90$+$19$-$5)/129) and the upper limit is $\sim$96\% 
(=(90$+$19$-$5)/(94$+$20$-$5)).

Secondly, we have investigated the completeness of the photometric
selection by examinining the characteristics of stars lying within 
the AAOmega field-of-view to which a fibre was assigned. After keeping 
only stars whose proper motion is within 3$\sigma$ from the mean proper 
motion of the cluster 
\citep[($-$11, $-$25) mas/yr;][]{deBruijne97,preibisch98}, we are left 
with 331 candidates.
As a second step we have removed giants and late-K/early-M dwarfs as 
well as M dwarfs which do not exhibit H$\alpha$ in emission, leaving
a total of 107 candidates. Then, we have kept only sources with a
Na {\small{I}} doublet equivalent width less than 5\,\AA{}, leaving 
58 candidates. We note that 10 (9 are members but 8 satisfied the
above criteria in the original selection) of these 58 sources lie 
to the red of the photometric selection and correspond to the proper motion 
candidates in Table \ref{tab_USco:PM_NM_L07a}. To summarise, the 
completeness of our photometric selection is better than 129/(129$+$58) 
$\sim$ 69\%, with an upper limit of 129/(129$+$8) $\sim$ 94\%.
These numbers are consistent with the values derived from our optical
and near-infrared spectroscopic follow-up as described in the first 
paragraph of this section.

Table \ref{tab_USco:missed_Memb} gives the coordinates (J2000), 
$ZYJHK$ photometry, proper motions (mas/yr), equivalent widths of 
the H$\alpha$ and Na {\small{I}} lines (in \AA{}), and spectral types 
for the 50 (58$-$8) candidates with proper motion within 3$\sigma$ from
the cluster mean motion and with equivalent widths measurements
of H$\alpha$ and the Na {\small{I}} doublet consistent with membership.
These objects are plotted as red open triangles in the ($Z-J$,$Z$) 
colour-magnitude diagram (Fig.\ \ref{fig_USco:CMD_ZJZ}).

Among the 58 sources left after the selection based on proper motions
and the equivalent widths of the H$\alpha$ line and the Na {\small{I}}
doublet, 11 exhibit lithium in absorption and 39 show noticeable 
N {\small{II}} emission at 6583\,\AA{}. Of the 39 with N {\small{II}} 
emission, four exhibit strong S {\small{II}} emission lines at 6716\,\AA{} 
and 6731\,\AA{}, indicative of a high excitation TTauri-like 
phase. In the full sample of stars in the AAOmega field-of-view, 87 sources
show N {\small{II}} in emission and 16 out of 87 exhibit strong 
S {\small{II}} emission lines.
They may belong to the nearby Upper Centaurus Lupus association as the
difference in the age (13--16 Myr vs 5 Myr) and the distance (145 pc vs 160 pc)
would result in members of Upper Centaurus Lupus being one magnitude
fainter than their USco counterparts according to theoretical models
\citep{baraffe98}. They could also have escaped from the nearby
$\rho$ Oph star-forming region ($\leq$ 1 Myr; 140 pc) where the occurence
of disks is expected to be higher than in Upper Sco \citep{haisch01}.

\subsection{Three young accreting sources}
\label{USco_AAOmega:Field_accreting}
%


Among stars in the AAOmega field followed-up spectroscopically, we have 
detected three accreting sources with strong emission lines. These sources, 
UGCS J161450.31$-$233240.0 (M4.5; Fig.\ \ref{fig_USco:spec_others}), 
UGCS J161503.64$-$235417.7 (M4), and UGCS J160729.59$-$230822.4 (M3), 
exhibit strong 
emission lines of H$\alpha$ and He {\small{I}} with pseudo equivalent 
widths of ($-$108,$-$3.0), ($-$187,$-$2.2), and ($-$150,$-$3.0)\,\AA{}, 
respectively. We also clearly detect the sodium doublet at 5875/5890\,\AA{}, 
the calcium triplet at 8498/8542/8662\,\AA{}, and O{\small{I}} forbidden
emission lines at $\sim$6300 and 6363, and 8446\,\AA{} in all three 
sources. The Ca triplet is weaker in
UGCS J161503.64$-$235417.7 than in the other two objects while the
O{\small{I}} emission is comparable with pseudo equivalent widths 
between 3.6 and 5.3\,\AA{}. The proper motions of these three 
sources are given in Table \ref{tab_USco:field_stars}: only 
UGCS J161503.64$-$235417.7 has a proper motion inconsistent with
the USco mean proper motion.
We have added to Table \ref{tab_USco:missed_Memb} the two accreting 
sources with proper motions consistent with USco for consistency
but we do not consider them as members in the rest of the paper
because they were originally classified as photometric non members.

The presence of a disk and strong emission lines indicates a young age
and a high excitation phase like in TTauri stars. The O{\small{I}} 
forbidden lines are indicative of outflows. These objects, however,
did not fall in our initial photometric selection because of their
colours. Indeed, young disk-bearing sources tends to exhibit unusual
colours and magnitudes leading to their rejection as photometric
candidates \citep[e.g.][]{luhman08c,lodieu09e,mayne10a}. A mid-infrared 
spectroscopic study of these two objects would shed light on the amount 
of dust re-processing \citep[e.g.][]{riaz09b} and lead to better constraints 
on the age of these sources in order to find out whether they are 
members of USco or of a younger population. Additional study
of the sources listed Table \ref{tab_USco:missed_Memb} should be
undertaken to find out if a pure photometric search is indeed biased
towards sources with low accretion rates \citep{mayne10a}.

%
%
\section{The spectroscopic mass function}
\label{USco_AAOmega:IMF}

In this section, we consider all photometric candidates confirmed
spectroscopically as members of USco. The original photometric sample
contains 129 sources \citep{lodieu07a}. After the optical spectroscopic 
follow-up conducted with AAOmega (90 members; this study) and the 
near-infrared cross-dispersed spectroscopy obtained for the faintest 
candidates \citep[19 members; 5 in common with optical follow-up][]{lodieu08a},
we have a final spectroscopic sample of 104 members. We should add to this 
sample the nine members originally rejected on astrometric grounds 
(Table \ref{tab_USco:PM_NM_L07a}), yielding a total of 113 members
in 6.5 square degrees near the centre of the USco association. 

Spectroscopy is missing for 20 photometric candidates. We have
looked at the histograms of the number of sources with and without
spectroscopy as a function of magnitude, dividing up each sample
into bins of 0.5 mag. The numbers of candidates without spectroscopy is 
of the order or below the square root of the number of objects with 
optical spectra in each magnitude bin. Therefore, we do not expect any
artificial gap or peak in a statistical point-of-view due to the
incomplete spectroscopic follow-up. However, this effect of incompleteness
is hard to quantify until spectroscopy is obtained for all photometric
candidates. Furthermore, most of the photometric candidates without 
spectroscopy are brighter than $J$ = 14 mag, corresponding to spectral 
types of M6.5 or earlier i.e. where the bulk of spectroscopic members 
is located. Thus, the influence on the peak of the luminosity function
should be minimal.

%
%
\begin{table*}
\caption{Number of sources as a function of spectral type
along with the assigned effective temperatures (in K) and
$J$-band bolometric corrections}
\label{tab_USco:SpT_Teff_BCj}
\centering
\begin{tabular}{@{\hspace{0mm}}c c c c c c c c c c c c c c c c c@{\hspace{0mm}}}
 \hline
 \hline
M3.5 & M4.0 & M4.5 & M5.0 & M5.5 & M5.75 & M6.25 & M6.5 & M7.0 & M7.5 & M8.0 & M8.5 & M9.0 & L0.0 & L1.0 & L2.0 \cr
 \hline
 3   &  7   &  20  &  25  &  6   &  22   &  5    &  3   &  2   &  3   &  0   &  3   &  1   &  6   &  6   &  1   \cr
3340 & 3270 & 3200 & 3125 & 3060 & 3025  & 2935  & 2910 & 2880 & 2800 & 2710 & 2550 & 2400 & 2250 & 2100 & 1950 \cr
1.85 & 1.86 & 1.86 & 1.90 & 2.02 & 2.03  & 2.04  & 2.04 & 2.04 & 2.02  & 2.01 & 2.00  & 1.97 & 1.92 & 1.92 & 1.91 \cr
 \hline
\end{tabular}
\end{table*}
\subsection{Effective temperatures}
\label{USco_AAOmega:HR}

The spectral types of these members from the spectroscopic
sample range from M3.25 to L2\@. The uncertainty is typically 0.25 subclass 
for spectral types earlier than M7.5 and 0.5 subclass for later types.

To estimate the temperature associated with each spectral type, we have
used the temperature scale defined by \citet{luhman99a} for young objects
with spectral types between M3 and M9 (Table \ref{tab_USco:SpT_Teff_BCj}). 
This scale uses temperatures intermediate between field dwarfs and giants 
and is valid up to M9\@. Typical uncertainties on each individual temperature
are of the order 50\,K\@. More recently, \citet{rice10} designed an
independent scale for late-M dwarfs relying on model fits to high-resolution 
near-infrared spectra. These authors adopted effective temperatures of
$\sim$2930\,K and 2850\,K for USco\,66AB (M6) and USco\,100 (M7), 
respectively, in agreement with the scale designed by \citet{luhman99a} 
within current error bars. For later spectral types, we kept the scale 
proposed by \citet{lodieu08a} i.e. 2250$\pm$50\,K for L0, 2100$\pm$100\,K 
for L1, and 1950$\pm$150\,K for L2\@. The temperatures adopted in this 
paper are compiled in Table \ref{tab_USco:SpT_Teff_BCj}.

We also defined a set of bolometric corrections in the $J$-band
\citep{luhman99a} for the full range of spectral types covered by our 
study. For late-M and early-L dwarfs, we considered the values
published by \citet{dahn02} and \citet{vrba04} whereas values for
earlier spectral types come from \citet{leggett00a}. The latter
bolometric correction are derived for old field dwarfs and may not 
be valid for younger brown dwarfs. The adopted mean values are compiled 
in the last row of Table \ref{tab_USco:SpT_Teff_BCj}.

%
%
%
\begin{figure}
   \centering
   \includegraphics[angle=0, width=\linewidth]{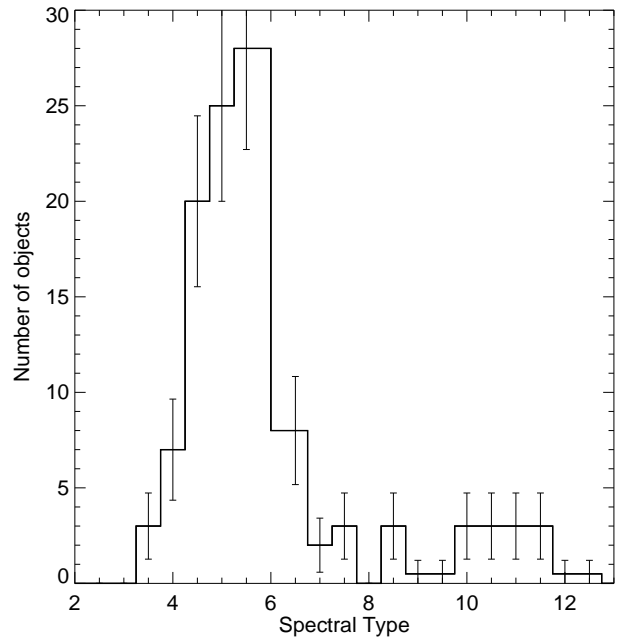}
   \caption{Luminosity function or number of USco spectroscopic members 
as a function of spectral types (4$\equiv$M4, 5$\equiv$M5, etc\ldots{}).
}
   \label{fig_USco:LF_hist}
\end{figure}
\subsection{The luminosity function}
\label{USco_AAOmega:LF}

For the sources in common between the optical and near-infrared spectroscopic 
follow-up, we have kept the spectral types derived from the optical spectra. 
Table \ref{tab_USco:SpT_Teff_BCj} lists the numbers of sources per
spectral type category. However, to plot the luminosity function,
we need to create bins of equal size which we choose to be 0.5 subtype. 
Therefore, we put together the sources classified as M5.5 and M5.75 
into the M5.5 bin, the objects with M6.25 and M6.5 into the M6.5 bin,
and divided the numbers of sources in the M9, L0, L1, and L2 categories
by two to split them up into 0.5 subtype bins. The resulting luminosity
function i.e.\ the number of objects as a function of spectral type and 
binned by half a subclass, is shown in Fig.\ \ref{fig_USco:LF_hist}.

Although our sample in the region studied here is complete, the spectral 
type of every object is associated with an error on the spectral assignment. 
Hence, we have plotted error bars that correspond to the square root of the 
count in each bin. The number of objects increases quickly towards M5, 
peaks around M6 and then decreases swiftly to M7 where the M7/M8 gap is 
proposed due to the onset of dust in substellar atmospheres 
\citep{dobbie02b}. The luminosity function appears flat towards later
spectral types and cooler temperatures. The overall trend is very similar 
to the analogous distribution for IC\,348 (valid for spectral types 
earlier than M9) depicted in Figure 11 of \citet{luhman03b}.

\begin{figure}
   \centering
   \includegraphics[angle=0, width=\linewidth]{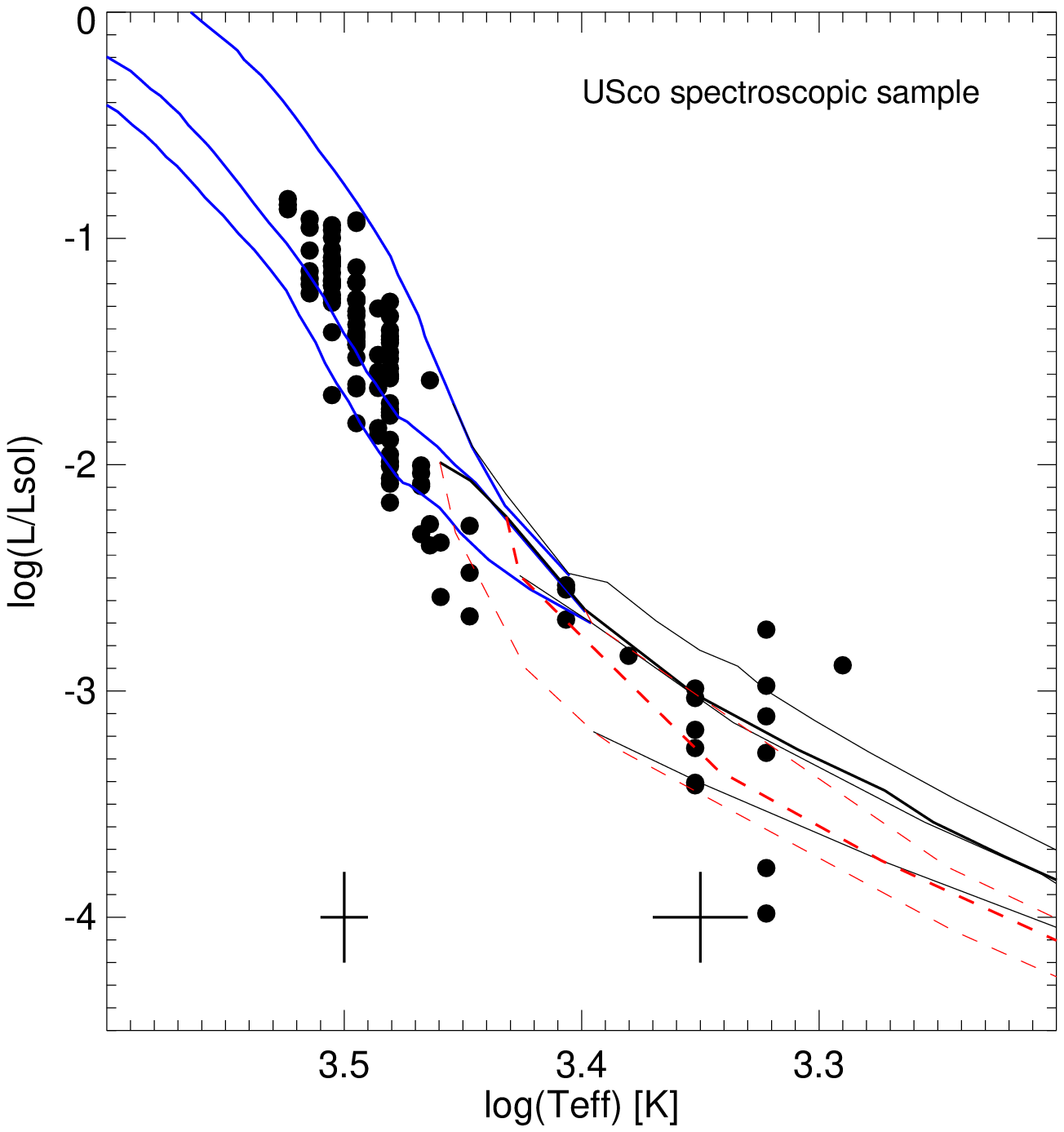}
   \caption{Hertzsprung-Russell diagram for USco spectroscopic
members confirmed at optical wavelengths with AAT/AAOmega in this paper 
and in the near-infrared with Gemini/GNIRS \citep{lodieu08a}.
Solid lines are NextGen and DUSTY isochrones at 1, 5, and 10 Myr.
Dashed lines represent isomasses of 0.05, 0.03, and
0.02 M$_{\odot}$ from the DUSTY models only \citep{chabrier00c}.
Typical error bars have been added at the bottom of the plot.
}
   \label{fig_USco:diag_HR}
\end{figure}

With the effective temperatures and the bolometric luminosities
derived for each individual object in our spectroscopic sample
of confirmed USco members, we can plot an Hertzsprung-Russell
diagram shown in Figure \ref{fig_USco:diag_HR}. Our diagram
is similar to Figure 8 of \citet{slesnick08} which involves 
a larger number of members due to the larger areal coverage. 
The mean age of our sample is around 5 Myr with a dispersion between 
1 and 10 Myr, consistent with the discussion presented in
\citet{slesnick08}. Our survey extends to lower effective temperatures 
and luminosities after including the brown dwarfs confirmed 
spectroscopically in the near-infrared with Gemini \citep{lodieu08a}.
Mixing optical and near-infrared spectral types may introduce
some biases at low masses and, thus, affect the determination of
the mass function. Indeed, near-infrared spectral types tend to be 
later than optical spectral types \citep{lodieu05b,luhman03b} by 
1--2 subclass (but not always; e.g.\ USco J160648.18$-$223040.1), yielding 
cooler temperatures by $\sim$100\,K and therefore lower masses.

\subsection{The mass function}
\label{USco_AAOmega:MF}

Several groups have developed models with different degrees of complexity to infer evolutionary tracks for young 
low-mass stars and brown dwarfs, e.g.\ \citet{palla93}, \citet{dantona94}, 
\citet{burrows97}, \citet{siess00}, \citet{baraffe98}, \citet{chabrier00c}, 
and \citep{baraffe02}. We refer 
the reader to the study by \citet{hillenbrand04} for a detailed
study of the discrepancies observed between models over a wide range 
of masses. For consistency with our earlier work in other regions 
targeted by the GCS \citep{lodieu07c,lodieu09e}, we have derived the 
bolometric luminosities using the apparent $J$ magnitudes and the 
bolometric corrections listed in Table \ref{tab_USco:SpT_Teff_BCj},
assuming  a bolometric magnitude of 4.74 for the Sun and a distance of 
145 pc for USco. Then to derive masses we have interpolated the luminosities 
provided
by the NextGen \citep{baraffe98} and DUSTY \citep{chabrier00c} for 
temperatures above and below 2500\,K, respectively.

\begin{figure*}
   \centering
   \includegraphics[angle=0, width=0.49\linewidth]{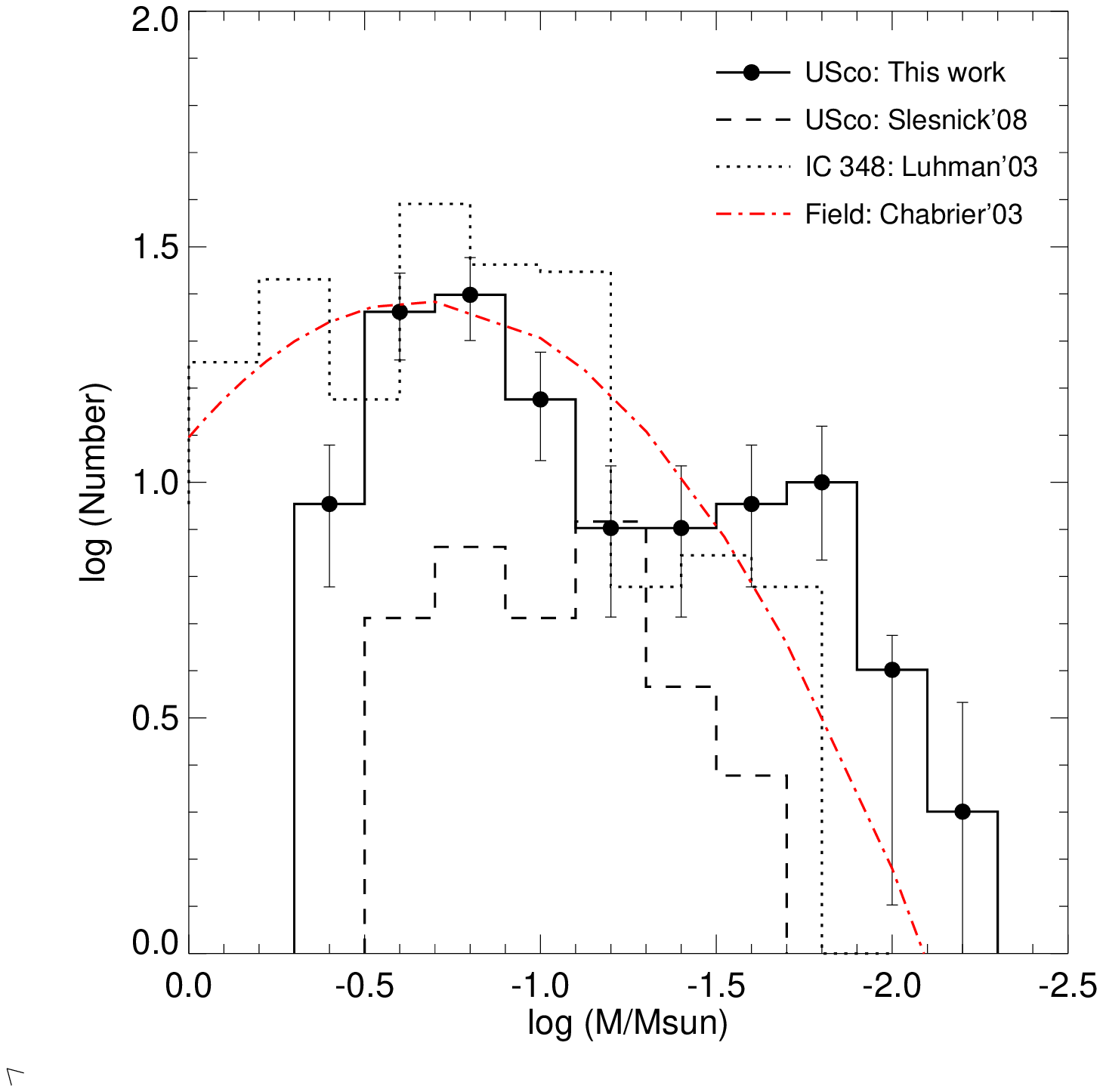}
   \includegraphics[angle=0, width=0.49\linewidth]{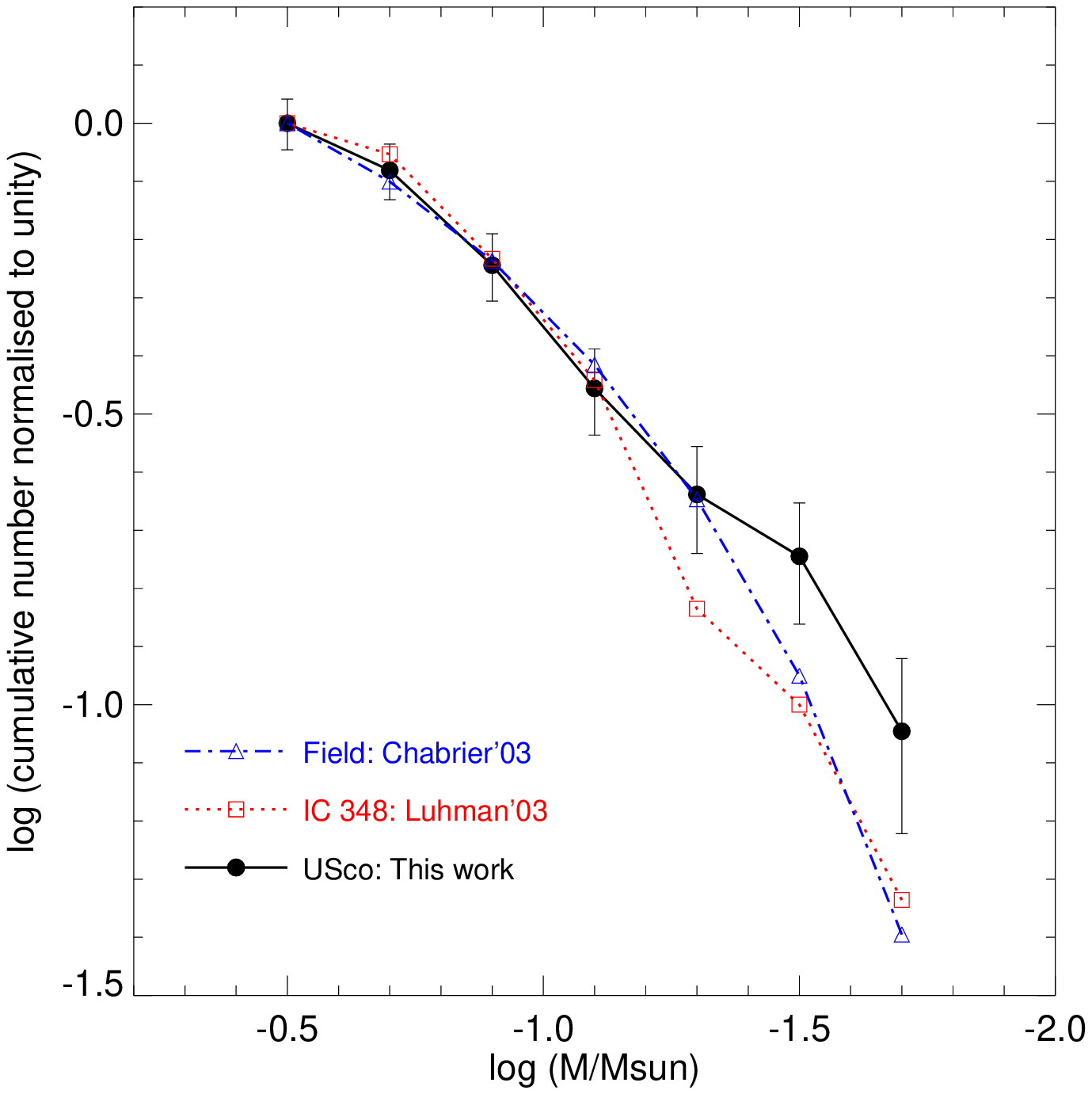}
   \caption{{\it{Left:}} Spectroscopic mass function for 6.5 square degree 
in the central region of USco valid between 0.35 and 0.01 M$_{\odot}$.
Overplotted as a dashed histogram is the spectroscopic mass function 
(corrected for incompleteness) published by \citet{slesnick08} and 
scaled to an area of 6.5 square degrees. The mass functions in IC\,348 
and the field are overplotted as a dotted histogram \citep{luhman03b} and 
dot-dash line \citep{chabrier03}, respectively.
{\it{Right:}} Cumulative distributions (logarithm of sum of the number 
of objects down to a given mass normalised to unity at the higher mass) for 
the mass functions in USco (filled dots and black line), IC\,348 (open 
squares and dotted line), and the field (open triangles and dot-dahed line)
over the mass range where the USco and IC\,348 mass functions are complete. 
Error bars are included for the USco mass function only.
}
   \label{fig_USco:MF_compare}
\end{figure*}

Our sample contains 113 spectroscopic members with spectral
types between M3.25 and L2 and masses between 0.4 and 0.006 M$_{\odot}$ 
identified in a 6.5 square degree area surveyed by the UKIDSS GCS\@. 
However, we estimate our photometric survey to be complete only down 
to 0.01 M$_{\odot}$. Only 20 objects of the original 
129 photometric member candidates \citep{lodieu07a} do not have 
spectra and are not included in the estimate of the mass function.
The resulting mass function is shown 
in Fig.\ \ref{fig_USco:MF_compare} along with its error bars (square 
root of the number of objects per mass bin) and compared to the 
spectroscopic mass function of \citet{slesnick08} created from a sample 
of 377 objects over the $\sim$0.6--0.02 M$_{\odot}$ mass range spread 
over 150 square degrees (dashed line in Fig.\ \ref{fig_USco:MF_compare}). 
However the spectroscopic follow-up presented in 
\citet{slesnick08} is based on a $\sim$15\% sampling, variable across 
the magnitude range, with a peak at around $r$ $\sim$ 19.5 mag and 
a sharp drop-off at fainter magnitudes. Therefore, we have
tentatively scaled their spectroscopic mass function (thick line
in Fig.\ 12 of \citet{slesnick08}) to our areal coverage 
(22.5 vs 6.5 square degrees represent a scaling factor of 3.46 in 
logarithmic units). We have also overplotted the mass function in  IC\,348 
\citep[dotted line in Fig.\ \ref{fig_USco:MF_compare};][]{luhman03b} without 
any scaling (the binning slightly differs from the USco mass function).

The first data point of our USco mass function is incomplete due to
saturation of the GCS photometric survey. Our mass functions seems to
peak at around 0.2$\pm$0.1 M$_{\odot}$ but this should be taken with
a grain of salt as our survey is incomplete at masses higher than
$\sim$0.4 M$_{\odot}$. If true however, the peak would be consistent with 
the characteristic mass found in the Pleiades \citep{moraux03,lodieu07c}, 
IC\,348, the Trapezium \citep{luhman00b}, and the field \citep{chabrier03}
mass functions. After a decline occuring around 0.1 M$_{\odot}$, our mass 
function flattens in the substellar regime. The decline seen beyond 
0.01 M$_{\odot}$ is due to the incompleteness of our photometric survey.

The overall shape of the USco mass function is similar to the mass 
function in IC\,348 derived by \citet{luhman03b} over the stellar mass range 
and beyond the hydrogen-burning limit as it can be seen from the mass 
function (left-hand side plot of Fig.\ \ref{fig_USco:MF_compare}) and in the 
cumulative distribution
(right-hand side plot of Fig.\ \ref{fig_USco:MF_compare}).
The number of objects found by \citet{slesnick08} in the stellar regime
is lower by a factor of two. One explanation for the discrepancy may be 
an underestimate of the incompleteness at the bright end of 
\citet{slesnick08}'s survey. The use of different theoretical isochrones 
to transform observables into masses is unlikely to account for such a 
large difference in the number of 0.3--0.1 M$_{\odot}$ stars. 
We also find a larger number of brown dwarfs below 0.03 M$_{\odot}$, 
indicating a poor sensitivity of \citet{slesnick08}'s optical survey to 
low-mass substellar objects. 

The cumulative distributions shown in the right-hand side plot of
Fig.\ \ref{fig_USco:MF_compare} represent the normalised number of objects 
down to a given mass in logarithmic units. All mass functions appear similar 
within the error bars over the 0.3--0.06 M$_{\odot}$ mass range as indicated 
by the mass functions. However, the cumulative mass functions plotted in the 
right-hand side panel of Fig.\ \ref{fig_USco:MF_compare} show that the number 
of brown dwarfs is higher (although possibly within current error bars)
in USco than IC\,348 and the extrapolation of the 
field mass function in the common region where the USco and IC\,348 surveys 
are complete. To quantify this hypothesis, we have computed the ratio of 
the number of stars (M = 0.35--0.08 M$_{\odot}$) and the number of brown 
dwarfs (M = 0.08--0.02 M$_{\odot}$) for IC\,348 and USco. We find ratios of 
21/104 = 0.20$\pm$0.03 for IC\,348 (\citet{luhman03b} finds 0.18$\pm$0.04 
for a slightly different definition of the star/brown dwarf ratio) and 
24/59 = 0.41$\pm$0.03 for USco, suggesting that USco may indeed contain an 
excess of substellar objects as originally suggested by \citet{preibisch02}
and corroborated by \citet{slesnick08}. As such, the USco association 
would be the best place to search for the turn down of the mass function 
and investigate the issue of the fragmentation limit.

%
%
%
%

%
%
%
%

%
%
\section{Conclusions}

We have presented a multi-fibre spectroscopic follow-up of more than
100 photometric candidates identified in the central part of the USco
association. The main results of our survey are:
 \begin{itemize}
 \item[$\bullet$] We have confirmed the spectroscopic membership of 90
photometric candidates based on the presence of H$\alpha$ in emission
and weak gravity-sensitive features in their optical spectra
 \item[$\bullet$] We have rejected four photometric candidates as members,
implying a completeness level of our original selection between 69 and 96\%
 \item[$\bullet$] $\sim$10\% of the new spectroscopic members show very
strong H$\alpha$ emission lines, indicating the presence of disks and
accretion
 \item[$\bullet$] We are able to detect lithium in absorption and the
magnesium triplet in most members despite only moderate spectral resolution
 \item[$\bullet$] We have demonstrated the efficiency of our original
photometric and proper motion selections
 \item[$\bullet$] Among the photometric non-members, we have identified 
three young disk-bearing sources showing signs of outflows
 \item[$\bullet$] We have assigned tentative spectral types to a large 
number of stars falling in the AAOmega field of view
 \item[$\bullet$] The luminosity drawn from a spectroscopic sample of 
113 sources with spectral types ranging from M3.25 to L2 shows a peak at 
around M6 and is flat at later spectral types
 \item[$\bullet$] The mass function derived in this 6.5 square degree
area in Upper Sco is very similar to that derived of IC\,348 and the 
extrapolation of the field mass function down to $\sim$0.06 M$_{\odot}$
 \item[$\bullet$] The mass function may peak around 0.2 M$_{\odot}$ and
is flat in the substellar regime
 \item[$\bullet$] The number of low mass brown dwarfs in USco appears higher
than in IC\,348 although within current error bars, pointing towards 
the original suggestion that USco might be rather rich in low-mass brown 
dwarfs and, thus, the best place to search for the turn down of the mass 
function
 \end{itemize}

The low level of contamination inferred from our original photometric and 
proper motion selection using data taken during the science verification 
phase of the UKIDSS GCS implies that similar selection can be applied
to the full association to carry out a global study of the region. 
The latest GCS data release includes $>$30\% coverage of the USco 
association. Subsequent data releases will soon complete the coverage in 
all filters and we will be able to identified new members to investigate
important issues, including the distribution of low-mass stars and
brown dwarfs with respect to massive stars and possible variation of 
the mass function within the association. The efficiency of the GCS
is favorable to future deep surveys of USco aiming at finding cooler
brown dwarfs and concentrating on the topic of the fragmentation limit.
The advent of VISTA will also provide a complementary second epoch in
USco to improve the astrometric selection.

Our work also has some implications for the study of other regions
surveyed by the GCS\@. The combination of two optical filters ($Z,Y$),
three infrared bands, and proper motion has confirmed the power of
the GCS in several regions, including the Pleiades, USco, and
$\sigma$ Orionis. Other regions such as Orion and Taurus are 
included in the GCS and will provide important clues to address
the issue of the universality of the IMF as a function of age and
environment.

%
%
\begin{acknowledgements}
NL acknowledges funding from the Spanish Ministry of Science and
Innovation through the Ram\'on y Cajal fellowship number 08-303-01-02\@. 
NL also thanks the Anglo-Australian Observatory (AAO) for funding part
of his stay at Epping in February/March 2010 and the AAO staff for an 
enjoyable visit.
Based on data obtained with the AAOmega spectrograph installed on the
Anglo Australian 3.9-m telescope in Siding Springs, Australia 
(program 07A/040). We thank Dr.\ Kevin Luhman for sending us the optical 
spectra of young M dwarfs and the data to plot the IMF of IC348\@.

The United Kingdom Infrared Telescope is operated by the Joint Astronomy 
Centre on behalf of the U.K.\ Science Technology and Facility Council.
This research has made use of the Simbad database, operated at
the Centre de Donn\'ees Astronomiques de Strasbourg (CDS), and
of NASA's Astrophysics Data System Bibliographic Services (ADS).
\end{acknowledgements}

%
%
  \bibliographystyle{aa}
  \bibliography{../mnemonic,../biblio_old}

%
%
\begin{appendix}
\section{Table New Members}
\longtab{5}{\scriptsize
\begin{landscape}
  \begin{longtable}{c c| c c c c| c c c c| c c c c| c c c c c c| c c c}
  \caption{\label{tab_USco:New_memb} Confirmed spectroscopic members
of the USco association AAOmega (first part) and GNIRS (second part).} \\
  \hline\hline
USco J\ldots{} & $J$ & \multicolumn{4}{c|}{EW(H$_{\alpha}$)} & \multicolumn{4}{c|}{EW(NaI)} & \multicolumn{4}{c|}{EW(KI)} & TiO7140 & TiO8465 & Na8189 & PC3 &  SpPC3 & SpT & T$_{\rm eff}$ & $\log$(L$_{\rm bol}$/L$_{\odot}$) & Mass \\
               & mag & \multicolumn{4}{c|}{\AA{}} & \multicolumn{4}{c|}{\AA{}} & \multicolumn{4}{c|}{\AA{}} &         &         &        &     &        &     &   Kelvins     &                                   & M$_{\odot}$ \\
\hline
\endfirsthead
\caption{continued.} \\
\hline\hline
USco J\ldots{} & $J$ & \multicolumn{4}{c|}{EW(H$_{\alpha}$)} & \multicolumn{4}{c|}{EW(NaI)} & \multicolumn{4}{c|}{EW(KI)} & TiO7140 & TiO8465 & Na8189 & PC3 &  SpPC3 & SpT & T$_{\rm eff}$ & L$_{\rm bol}$/L$_{\odot}$ & Mass \\
               & mag & \multicolumn{4}{c|}{\AA{}} & \multicolumn{4}{c|}{\AA{}} & \multicolumn{4}{c|}{\AA{}} &         &         &        &     &        &     &   Kelvins     &                                   & M$_{\odot}$ \\
\hline
\endhead
\hline
\endfoot
16:08:05.54$-$22:18:07.1 & 10.762 &    $-$4.4 &    $-$4.5 &     --- &     --- & 3.6 & 3.4 & --- & --- & 1.6 & 1.5 & --- & --- & 1.40 & 1.12 & 0.90 & 0.93 & 2.5 & 3.25 & 3340 & $-$0.826 & 0.4100 \\
16:08:15.66$-$22:22:20.1 & 10.825 &    $-$3.9 &    $-$3.8 &     --- &     --- & 3.8 & 3.7 & --- & --- & 1.6 & 1.9 & --- & --- & 1.43 & 1.10 & 0.91 & 0.94 & 2.6 & 3.25 & 3340 & $-$0.851 & 0.3937 \\
16:08:23.03$-$23:35:29.5 & 10.875 &    $-$5.8 &    $-$5.6 &    $-$5.6 &     --- & 3.6 & 3.4 & 3.5 & --- & 1.7 & 1.7 & 1.6 & --- & 1.42 & 1.12 & 0.89 & 0.93 & 2.5 & 3.25 & 3340 & $-$0.871 & 0.3826 \\
16:10:01.84$-$23:49:43.4 & 10.949 &   $-$13.1 &   $-$10.2 &   $-$10.6 &     --- & 3.5 & 3.3 & 3.4 & --- & 1.8 & 2.1 & 2.1 & --- & 1.94 & 1.24 & 0.89 & 1.00 & 3.0 &  5.0 & 3125 & $-$0.921 & 0.3551 \\
16:09:02.01$-$23:22:40.3 & 10.973 &    $-$4.6 &    $-$4.6 &    $-$4.5 &     --- & 3.4 & 3.4 & 3.5 & --- & 2.5 & 1.4 & 1.4 & --- & 1.64 & 1.15 & 0.90 & 0.93 & 2.4 &  4.0 & 3270 & $-$0.914 & 0.3586 \\
16:11:37.84$-$22:10:27.5 & 11.069 &    $-$3.1 &    $-$3.0 &    $-$3.0 &     --- & 3.4 & 2.9 & 3.1 & --- & 1.3 & 1.3 & --- & --- & 1.58 & 1.12 & 0.91 & 0.92 & 2.4 &  4.0 & 3270 & $-$0.953 & 0.3373 \\
16:13:54.34$-$23:20:34.4 & 11.095 &  $-$156.6 &  $-$155.5 &  $-$137.5 &     --- & 3.6 & 3.3 & 3.3 & --- & 1.4 & 1.6 & 1.4 & --- & 1.68 & 1.22 & 0.90 & 0.95 & 2.6 &  4.5 & 3200 & $-$0.963 & 0.3315 \\
16:10:23.44$-$23:12:17.7 & 11.041 &    $-$9.2 &    $-$8.2 &    $-$8.5 &     --- & 4.2 & 3.7 & 4.0 & --- & 2.0 & 2.0 & 2.0 & --- & 1.89 & 1.22 & 0.88 & 0.93 & 2.5 &  4.5 & 3200 & $-$0.942 & 0.3435 \\
16:15:54.87$-$23:15:14.8 & 11.180 &     --- &     --- &    $-$6.2 &     --- & --- & --- & 4.0 & --- & --- & --- & 1.8 & --- & 1.76 & 1.18 & 0.88 & 0.95 & 2.6 &  4.5 & 3200 & $-$0.997 & 0.3126 \\
16:15:27.43$-$22:39:27.7 & 11.321 &     --- &     --- &    $-$4.3 &     --- & --- & --- & 3.7 & --- & --- & --- & 1.6 & --- & 1.64 & 1.13 & 0.89 & 0.92 & 2.4 &  4.0 & 3270 & $-$1.054 & 0.2847 \\
16:12:39.54$-$22:28:08.3 & 11.308 &    $-$4.3 &    $-$4.3 &    $-$4.2 &     --- & 3.4 & 3.2 & 3.4 & --- & 1.4 & 1.4 & --- & --- & 1.69 & 1.16 & 0.90 & 0.94 & 2.5 &  4.5 & 3200 & $-$1.048 & 0.2871 \\
16:09:09.39$-$22:45:59.1 & 11.398 &    $-$6.8 &    $-$6.8 &    $-$7.2 &     --- & 4.1 & 3.8 & 3.8 & --- & 1.8 & 1.9 & 1.7 & --- & 1.61 & 1.14 & 0.88 & 0.94 & 2.5 &  4.5 & 3200 & $-$1.084 & 0.2707 \\
16:10:26.50$-$22:30:53.4 & 11.467 &    $-$8.4 &    $-$9.7 &    $-$9.0 &     --- & 3.4 & 3.3 & 3.8 & --- & 1.8 & --- & 1.5 & --- & 1.92 & 1.24 & 0.89 & 0.99 & 2.9 &  5.0 & 3125 & $-$1.128 & 0.2509 \\
16:14:38.46$-$23:21:37.3 & 11.436 &     --- &     --- &    $-$2.8 &     --- & --- & --- & 3.5 & --- & --- & --- & 1.3 & --- & 1.53 & 1.10 & 0.90 & 0.95 & 2.6 &  4.5 & 3200 & $-$1.100 & 0.2638 \\
16:15:42.06$-$22:35:24.0 & 11.436 &     --- &     --- &    $-$5.4 &     --- & --- & --- & 3.1 & --- & --- & --- & 1.4 & --- & 1.66 & 1.14 & 0.90 & 0.91 & 2.3 &  4.5 & 3200 & $-$1.100 & 0.2638 \\
16:09:58.78$-$23:54:27.5 & 11.490 &    $-$4.6 &    $-$5.0 &    $-$4.3 &     --- & 3.4 & 3.4 & 3.1 & --- & 1.7 & 1.6 & 1.6 & --- & 1.83 & 1.18 & 0.89 & 0.95 & 2.6 &  4.5 & 3200 & $-$1.121 & 0.2540 \\
16:11:23.99$-$22:53:32.6 & 11.549 &    $-$5.2 &    $-$5.0 &    $-$4.7 &     --- & 3.7 & 3.5 & 3.5 & --- & --- & 1.8 & 1.9 & --- & 1.54 & 1.11 & 0.89 & 0.94 & 2.5 &  4.0 & 3270 & $-$1.145 & 0.2443 \\
16:08:08.47$-$22:25:00.1 & 11.631 &    $-$6.8 &    $-$6.5 &     --- &     --- & 4.2 & 3.8 & --- & --- & 1.9 & 1.8 & --- & --- & 1.56 & 1.13 & 0.88 & 0.95 & 2.6 &  4.0 & 3270 & $-$1.178 & 0.2317 \\
16:07:06.33$-$22:48:28.2 & 11.567 &    $-$5.7 &     --- &     --- &     --- & 3.5 & --- & --- & --- & 1.7 & --- & --- & --- & 1.71 & 1.17 & 0.90 & 1.00 & 3.0 &  4.5 & 3200 & $-$1.152 & 0.2415 \\
16:14:41.19$-$22:27:05.4 & 11.664 &     --- &     --- &   $-$47.0 &     --- & --- & --- & 3.1 & --- & --- & --- & 1.5 & --- & 1.87 & 1.20 & 0.89 & 0.94 & 2.5 &  4.5 & 3200 & $-$1.191 & 0.2266 \\
16:11:17.05$-$22:13:08.8 & 11.639 &    $-$8.1 &    $-$7.0 &    $-$6.7 &     --- & 3.4 & 3.3 & 3.2 & --- & 1.7 & 1.6 & --- & --- & 1.95 & 1.24 & 0.90 & 0.98 & 2.8 &  5.0 & 3125 & $-$1.197 & 0.2243 \\
16:09:39.68$-$22:31:53.9 & 11.630 &    $-$5.7 &    $-$4.6 &    $-$4.8 &     --- & 3.3 & 3.3 & 3.0 & --- & --- & --- & 1.4 & --- & 1.88 & 1.20 & 0.90 & 0.98 & 2.8 &  5.0 & 3125 & $-$1.193 & 0.2257 \\
16:11:19.07$-$23:19:20.4 & 11.635 &    $-$3.5 &    $-$4.0 &    $-$3.6 &     --- & 3.4 & 3.4 & 3.0 & --- & 1.8 & 1.9 & 1.9 & --- & 1.84 & 1.18 & 0.90 & 1.01 & 3.1 &  5.0 & 3125 & $-$1.195 & 0.2249 \\
16:08:14.00$-$22:47:39.4 & 11.649 &    $-$8.3 &    $-$7.3 &     --- &     --- & 3.7 & 3.3 & --- & --- & 1.8 & 1.7 & --- & --- & 1.86 & 1.21 & 0.89 & 0.95 & 2.6 &  4.5 & 3200 & $-$1.185 & 0.2289 \\
16:14:57.50$-$23:28:42.8 & 11.695 &     --- &     --- &    $-$6.8 &     --- & --- & --- & 3.7 & --- & --- & --- & 1.7 & --- & 1.64 & 1.11 & 0.88 & 0.92 & 2.4 &  4.0 & 3270 & $-$1.203 & 0.2218 \\
16:08:50.34$-$22:03:28.7 & 11.709 &    $-$8.5 &    $-$8.3 &     --- &     --- & 3.8 & 3.7 & --- & --- & 1.6 & 1.9 & --- & --- & 1.83 & 1.20 & 0.88 & 0.97 & 2.8 &  4.5 & 3200 & $-$1.209 & 0.2197 \\
16:13:12.16$-$23:15:16.6 & 11.702 &    $-$5.4 &    $-$6.4 &    $-$5.6 &     --- & 3.5 & 3.3 & 3.4 & --- & 1.8 & 1.8 & 1.9 & --- & 1.84 & 1.20 & 0.89 & 0.95 & 2.6 &  4.5 & 3200 & $-$1.206 & 0.2207 \\
16:08:46.06$-$22:46:59.4 & 11.794 &    $-$8.7 &    $-$8.2 &    $-$9.4 &     --- & 4.2 & 3.9 & 3.9 & --- & 2.2 & 2.1 & 1.9 & --- & 1.64 & 1.15 & 0.87 & 0.95 & 2.6 &  4.0 & 3270 & $-$1.243 & 0.2066 \\
16:14:23.12$-$22:19:33.9 & 11.717 &     --- &     --- &   $-$94.5 &     --- & --- & --- & 3.4 & --- & --- & --- & 2.4 & --- & 2.09 & 1.46 & 0.89 & 1.05 & 3.4 & 5.75 & 3025 & $-$1.280 & 0.1928 \\
16:13:36.88$-$23:27:29.9 & 11.812 &     --- &    $-$5.5 &     --- &     --- & --- & 3.6 & --- & --- & --- & 1.8 & --- & --- & 1.79 & 1.17 & 0.88 & 0.93 & 2.5 &  4.5 & 3200 & $-$1.250 & 0.2038 \\
16:10:20.87$-$23:31:55.7 & 11.801 &   $-$15.1 &   $-$15.7 &   $-$14.3 &     --- & 3.6 & 3.5 & 3.6 & --- & 2.8 & 2.5 & 2.5 & --- & 2.41 & 1.46 & 0.88 & 1.02 & 3.1 &  5.5 & 3060 & $-$1.310 & 0.1823 \\
16:06:34.61$-$22:55:04.4 & 11.835 &   $-$12.7 &   $-$10.0 &     --- &     --- & 3.6 & 3.1 & --- & --- & 1.5 & 1.6 & --- & --- & 2.06 & 1.33 & 0.90 & 1.03 & 3.2 &  5.0 & 3125 & $-$1.275 & 0.1946 \\
16:11:02.11$-$23:35:50.6 & 11.817 &   $-$11.6 &   $-$12.8 &   $-$12.2 &     --- & 3.7 & 3.2 & 3.6 & --- & 2.0 & 1.9 & 2.0 & --- & 2.08 & 1.29 & 0.89 & 1.00 & 3.0 &  5.0 & 3125 & $-$1.268 & 0.1971 \\
16:06:50.18$-$23:09:54.0 & 11.816 &    $-$8.4 &    $-$9.0 &     --- &     --- & 4.1 & 3.4 & --- & --- & 2.1 & 2.1 & --- & --- & 1.91 & 1.29 & 0.88 & 1.06 & 3.4 &  5.0 & 3125 & $-$1.268 & 0.1973 \\
16:12:43.74$-$23:08:23.2 & 11.836 &    $-$5.9 &    $-$6.6 &    $-$5.6 &     --- & 3.2 & 3.4 & 3.1 & --- & 1.7 & 1.8 & 1.6 & --- & 2.00 & 1.25 & 0.89 & 0.99 & 3.0 &  5.0 & 3125 & $-$1.276 & 0.1944 \\
16:16:41.63$-$23:15:39.1 & 11.854 &     --- &     --- &    $-$5.4 &     --- & --- & --- & 3.7 & --- & --- & --- & 1.9 & --- & 1.81 & 1.17 & 0.88 & 0.94 & 2.5 &  4.5 & 3200 & $-$1.267 & 0.1976 \\
16:16:11.72$-$23:27:05.2 & 11.898 &     --- &     --- &    $-$5.7 &     --- & --- & --- & 3.4 & --- & --- & --- & 1.8 & --- & 1.85 & 1.19 & 0.89 & 0.95 & 2.6 &  4.5 & 3200 & $-$1.284 & 0.1913 \\
16:09:16.89$-$23:41:32.6 & 11.876 &   $-$12.0 &   $-$12.2 &   $-$13.1 &     --- & 3.6 & 3.4 & 3.4 & --- & 2.0 & 1.8 & 1.7 & --- & 2.17 & 1.33 & 0.89 & 1.05 & 3.4 & 5.75 & 3025 & $-$1.344 & 0.1712 \\
16:09:46.33$-$22:55:33.6 & 11.998 &   $-$10.3 &    $-$7.8 &    $-$8.5 &     --- & 4.3 & 3.9 & 3.7 & --- & --- & 2.0 & 1.9 & --- & 1.95 & 1.24 & 0.88 & 1.00 & 3.0 &  5.0 & 3125 & $-$1.340 & 0.1721 \\
16:13:41.30$-$23:54:22.0 & 11.951 &     --- &     --- &    $-$5.7 &     --- & --- & --- & 3.6 & --- & --- & --- & 2.1 & --- & 1.90 & 1.20 & 0.89 & 1.00 & 3.0 &  5.0 & 3125 & $-$1.322 & 0.1780 \\
16:11:37.61$-$23:46:14.8 & 12.099 &    $-$6.8 &    $-$6.1 &    $-$6.3 &     --- & 3.4 & 3.4 & 3.3 & --- & 1.8 & 1.6 & 1.7 & --- & 1.99 & 1.24 & 0.89 & 1.00 & 3.0 &  5.0 & 3125 & $-$1.381 & 0.1609 \\
16:13:36.47$-$23:27:35.5 & 12.098 &   $-$10.2 &     --- &   $-$11.8 &     --- & 4.0 & --- & 4.0 & --- & 2.9 & --- & 3.1 & --- & 2.21 & 1.35 & 0.86 & 1.03 & 3.2 & 5.75 & 3025 & $-$1.432 & 0.1464 \\
16:08:02.17$-$22:59:05.9 & 12.135 &   $-$20.2 &   $-$15.2 &   $-$17.7 &     --- & 3.6 & 3.7 & 3.6 & --- & 3.2 & 3.0 & 3.2 & --- & 2.17 & 1.38 & 0.87 & 1.05 & 3.4 & 5.75 & 3025 & $-$1.447 & 0.1422 \\
16:15:08.92$-$23:45:04.9 & 12.224 &     --- &     --- &    $-$7.5 &     --- & --- & --- & 4.0 & --- & --- & --- & 2.8 & --- & 1.83 & 1.18 & 0.86 & 0.96 & 2.7 &  4.5 & 3200 & $-$1.415 & 0.1514 \\
16:15:59.26$-$23:29:36.5 & 12.221 &     --- &     --- &    $-$7.1 &     --- & --- & --- & 3.6 & --- & --- & --- & 1.8 & --- & 1.95 & 1.21 & 0.89 & 0.97 & 2.8 &  5.0 & 3125 & $-$1.430 & 0.1472 \\
16:16:33.43$-$23:27:21.2 & 12.195 &     --- &     --- &   $-$10.6 &     --- & --- & --- & 3.6 & --- & --- & --- & 2.8 & --- & 2.02 & 1.32 & 0.86 & 1.03 & 3.2 &  5.0 & 3125 & $-$1.419 & 0.1502 \\
16:11:07.38$-$22:28:50.3 & 12.171 &   $-$64.7 &   $-$59.8 &   $-$61.3 &     --- & 3.4 & 3.3 & 3.4 & --- & 2.7 & 2.8 & --- & --- & 2.28 & 1.59 & 0.89 & 1.12 & 3.9 & 5.75 & 3025 & $-$1.462 & 0.1381 \\
16:14:13.52$-$22:44:58.0 & 12.283 &    $-$6.5 &    $-$5.8 &    $-$6.2 &     --- & 3.5 & 3.1 & 3.5 & --- & 1.9 & 1.7 & 1.6 & --- & 1.99 & 1.20 & 0.91 & 0.99 & 3.0 &  5.0 & 3125 & $-$1.454 & 0.1402 \\
16:07:50.39$-$22:21:02.2 & 12.277 &   $-$14.3 &   $-$12.5 &     --- &     --- & 3.8 & 3.5 & --- & --- & 2.2 & 2.3 & --- & --- & 2.22 & 1.38 & 0.89 & 1.07 & 3.6 & 5.75 & 3025 & $-$1.504 & 0.1272 \\
16:06:39.22$-$22:48:34.2 & 12.322 &    $-$6.9 &    $-$5.4 &    $-$7.1 &     --- & 3.4 & 3.5 & 3.3 & --- & --- & 1.7 & 2.1 & --- & 1.95 & 1.27 & 0.89 & 1.03 & 3.3 &  5.0 & 3125 & $-$1.470 & 0.1357 \\
16:12:14.92$-$22:18:04.0 & 12.320 &    $-$4.5 &    $-$5.4 &    $-$5.2 &     --- & 3.4 & 3.3 & 3.9 & --- & 1.7 & 2.0 & --- & --- & 1.98 & 1.23 & 0.89 & 0.98 & 2.9 &  5.0 & 3125 & $-$1.469 & 0.1359 \\
16:12:55.28$-$22:26:54.4 & 12.314 &    $-$6.8 &    $-$6.1 &    $-$6.7 &     --- & 3.5 & 3.0 & 3.6 & --- & 1.7 & 1.9 & --- & --- & 2.11 & 1.30 & 0.89 & 1.00 & 3.0 &  5.5 & 3060 & $-$1.515 & 0.1250 \\
16:07:45.21$-$22:22:57.6 & 12.348 &   $-$10.4 &    $-$9.6 &     --- &     --- & 4.0 & 3.8 & --- & --- & 2.5 & 2.3 & --- & --- & 2.19 & 1.37 & 0.88 & 1.10 & 3.8 & 5.75 & 3025 & $-$1.532 & 0.1215 \\
16:13:10.82$-$23:13:51.6 & 12.454 &   $-$11.5 &   $-$11.1 &   $-$10.0 &     --- & 4.7 & 4.8 & 4.2 & --- & 4.8 & 4.0 & 3.2 & --- & 2.25 & 1.35 & 0.83 & 1.02 & 3.2 & 5.75 & 3025 & $-$1.575 & 0.1130 \\
16:08:48.37$-$23:41:21.0 & 12.463 &    $-$9.4 &    $-$6.8 &   $-$10.8 &     --- & 3.5 & 3.0 & 4.7 & --- & 2.0 & 1.9 & 2.3 & --- & 1.94 & 1.36 & 0.89 & 1.05 & 3.4 &  5.0 & 3125 & $-$1.526 & 0.1227 \\
16:11:40.40$-$23:11:34.9 & 12.525 &   $-$10.9 &   $-$11.1 &   $-$10.8 &     --- & 4.5 & 3.8 & 4.4 & --- & 3.6 & 2.9 & 3.5 & --- & 2.27 & 1.39 & 0.85 & 1.06 & 3.5 & 5.75 & 3025 & $-$1.603 & 0.1073 \\
16:15:38.66$-$22:40:37.3 & 12.562 &     --- &     --- &   $-$10.6 &     --- & --- & --- & 3.8 & --- & --- & --- & 2.5 & --- & 2.14 & 1.31 & 0.87 & 1.02 & 3.1 & 5.75 & 3025 & $-$1.618 & 0.1044 \\
16:08:10.81$-$22:29:42.9 & 12.542 &   $-$22.0 &   $-$21.6 &     --- &     --- & 3.8 & 3.3 & --- & --- & 2.5 & 2.4 & --- & --- & 1.94 & 1.45 & 0.88 & 1.08 & 3.6 & 5.75 & 3025 & $-$1.610 & 0.1060 \\
16:09:58.52$-$23:45:18.7 & 12.574 &   $-$34.4 &   $-$37.6 &   $-$34.8 &     --- & 3.0 & 3.4 & 3.6 & --- & --- & 4.3 & 3.5 & --- & 2.77 & 1.74 & 0.89 & 1.26 & 4.9 &  6.5 & 2935 & $-$1.627 & 0.1026 \\
16:11:31.81$-$22:37:08.3 & 12.679 &    $-$6.8 &    $-$6.3 &    $-$5.9 &     --- & 3.5 & 3.1 & 3.0 & --- & 2.1 & 1.5 & 2.1 & --- & 2.07 & 1.37 & 0.90 & 1.05 & 3.4 &  5.5 & 3060 & $-$1.661 & 0.0965 \\
16:16:11.83$-$23:16:26.9 & 12.758 &     --- &     --- &    $-$8.3 &     --- & --- & --- & 3.9 & --- & --- & --- & 2.5 & --- & 1.99 & 1.25 & 0.87 & 0.99 & 3.0 &  5.0 & 3125 & $-$1.644 & 0.0993 \\
16:10:57.28$-$23:59:54.1 & 12.803 &    $-$7.1 &    $-$5.5 &    $-$4.7 &     --- & 3.4 & 3.7 & 3.1 & 2.5 & 1.9 & 2.2 & --- & --- & 2.06 & 1.26 & 0.88 & 0.99 & 2.9 &  5.0 & 3125 & $-$1.662 & 0.0963 \\
16:15:20.10$-$23:33:54.7 & 12.838 &     --- &     --- &   $-$12.1 &     --- & --- & --- & 3.5 & --- & --- & --- & 3.0 & --- & 2.34 & 1.47 & 0.85 & 1.11 & 3.9 & 5.75 & 3025 & $-$1.728 & 0.0853 \\
16:10:54.29$-$23:09:11.1 & 12.975 &    $-$8.2 &    $-$8.9 &    $-$9.4 &     --- & 5.1 & 5.2 & 4.6 & --- & 4.3 & 4.5 & 4.4 & --- & 2.24 & 1.35 & 0.83 & 1.03 & 3.3 & 5.75 & 3025 & $-$1.783 & 0.0761 \\
16:08:22.29$-$22:17:03.0 & 12.939 &   $-$18.9 &   $-$12.5 &     --- &     --- & 3.6 & 3.4 & --- & --- & 1.9 & 2.1 & --- & --- & 2.13 & 1.44 & 0.88 & 1.08 & 3.6 & 5.75 & 3025 & $-$1.769 & 0.0785 \\
16:09:04.51$-$22:24:52.5 & 12.918 &    $-$7.7 &   $-$11.3 &     --- &     --- & 4.7 & 4.0 & --- & --- & 2.6 & --- & --- & --- & 1.77 & 1.16 & 0.82 & 0.93 & 2.5 &  4.5 & 3200 & $-$1.692 & 0.0913 \\
16:15:28.19$-$23:15:44.1 & 13.123 &     --- &     --- &    $-$6.3 &     --- & --- & --- & 4.7 & --- & --- & --- & 3.0 & --- & 2.05 & 1.26 & 0.85 & 1.04 & 3.3 &  5.5 & 3060 & $-$1.838 & 0.0691 \\
16:12:16.09$-$23:44:25.0 & 13.188 &     --- &     --- &    $-$7.2 &    $-$4.9 & 4.4 & 3.9 & 4.0 & 3.2 & 2.6 & 2.3 & 2.9 & 2.0 & 1.90 & 1.29 & 0.86 & 0.99 & 2.9 &  5.0 & 3125 & $-$1.816 & 0.0714 \\
16:06:26.37$-$23:06:11.4 & 13.205 &   $-$14.4 &   $-$10.0 &     --- &     --- & 4.1 & 3.2 & --- & --- & 2.2 & 2.2 & --- & --- & 1.86 & 1.29 & 0.88 & 1.06 & 3.5 &  5.5 & 3060 & $-$1.871 & 0.0654 \\
16:11:34.70$-$22:19:44.3 & 13.242 &    $-$7.1 &   $-$33.3 &   $-$30.0 &   $-$19.4 & 3.7 & 3.5 & 3.0 & 3.0 & 2.4 & 2.3 & 2.7 & 2.4 & 2.20 & 1.35 & 0.88 & 1.06 & 3.5 & 5.75 & 3025 & $-$1.890 & 0.0633 \\
16:11:26.30$-$23:40:06.1 & 13.404 &     --- &     --- &    $-$7.4 &    $-$7.9 & 6.0 & 5.2 & 3.7 & 3.1 & --- & 3.1 & 2.8 & 2.0 & 2.26 & 1.35 & 0.88 & 1.07 & 3.5 & 5.75 & 3025 & $-$1.955 & 0.0556 \\
16:13:34.76$-$23:28:15.7 & 13.485 &    $-$9.7 &   $-$11.2 &   $-$10.3 &    $-$8.7 & 3.5 & 3.5 & 4.0 & 3.5 & 2.3 & 3.0 & 2.7 & 3.1 & 2.17 & 1.33 & 0.86 & 1.01 & 3.1 & 5.75 & 3025 & $-$1.987 & 0.0516 \\
16:13:26.66$-$22:30:35.0 & 13.515 &     --- &    $-$8.2 &     --- &     --- & --- & 3.1 & --- & --- & 3.1 & 3.9 & --- & --- & 2.54 & 1.65 & 0.88 & 1.12 & 3.9 & 6.25 & 2910 & $-$2.003 & 0.0496 \\
16:12:46.92$-$23:38:40.9 & 13.601 &   $-$17.8 &   $-$12.7 &   $-$13.8 &   $-$17.4 & 3.7 & 3.9 & 4.4 & 3.3 & 3.9 & 4.8 & 3.8 & --- & 2.59 & 1.61 & 0.86 & 1.16 & 4.2 & 6.25 & 2910 & $-$2.038 & 0.0453 \\
16:11:57.37$-$22:15:06.8 & 13.668 &    $-$8.1 &    $-$8.9 &    $-$9.2 &    $-$8.8 & 3.4 & 3.3 & 2.8 & 3.7 & 2.4 & 3.0 & --- & 2.9 & 2.16 & 1.35 & 0.87 & 1.06 & 3.5 & 5.75 & 3025 & $-$2.060 & 0.0424 \\
16:16:45.39$-$23:33:41.6 & 13.726 &     --- &     --- &    $-$9.5 &    $-$7.6 & --- & --- & 4.7 & 3.7 & --- & --- & 3.4 & 3.4 & 2.28 & 1.38 & 0.85 & 1.06 & 3.5 & 5.75 & 3025 & $-$2.084 & 0.0398 \\
16:13:42.64$-$23:01:28.0 & 13.718 &    $-$9.5 &    $-$9.7 &   $-$10.8 &    $-$7.3 & 4.7 & 3.9 & 4.4 & 3.6 & 2.9 & 4.2 & 3.5 & 4.1 & 2.50 & 1.51 & 0.85 & 1.15 & 4.1 & 6.25 & 2910 & $-$2.084 & 0.0397 \\
16:11:38.37$-$23:07:07.5 & 13.743 &   $-$14.9 &    $-$7.8 &   $-$11.9 &    $-$9.4 & 3.4 & 3.3 & 3.8 & 3.1 & 2.7 & 3.0 & 2.8 & 2.4 & 2.37 & 1.51 & 0.87 & 1.12 & 3.9 & 6.25 & 2910 & $-$2.094 & 0.0391 \\
16:15:36.48$-$23:15:17.6 & 13.935 &     --- &     --- &   $-$11.8 &   $-$12.0 & --- & --- & 4.2 & 3.9 & --- & --- & 3.6 & 3.7 & 2.28 & 1.38 & 0.85 & 1.08 & 3.6 & 5.75 & 3025 & $-$2.167 & 0.0349 \\
16:14:32.87$-$22:42:13.5 & 14.163 &   $-$73.4 &   $-$88.1 &   $-$59.8 &   $-$64.8 & 3.7 & 2.9 & 3.3 & 2.4 & --- & --- & 3.7 & 2.5 & 2.38 & 1.85 & 0.90 & 1.26 & 4.8 &  6.5 & 2935 & $-$2.262 & 0.0297 \\
16:11:54.39$-$22:36:49.3 & 14.274 &   $-$10.3 &    $-$9.3 &   $-$14.5 &   $-$16.8 & 4.2 & 3.7 & 4.1 & 3.7 & 5.5 & 4.0 & 5.0 & 3.9 & 2.60 & 1.53 & 0.85 & 1.14 & 4.0 & 6.25 & 2910 & $-$2.307 & 0.0286 \\
16:09:29.39$-$23:43:12.2 & 14.202 &    $-$8.9 &   $-$18.9 &     --- &     --- & 4.4 & 3.1 & --- & --- & 6.9 & 5.8 & --- & --- & 2.72 & 1.95 & 0.83 & 1.29 & 5.0 &  7.5 & 2800 & $-$2.270 & 0.0295 \\
16:08:07.45$-$23:45:05.6 & 14.398 &   $-$81.0 &   $-$56.6 &     --- &     --- & 3.6 & 3.6 & --- & --- & --- & 3.4 & --- & --- & 2.11 & 1.71 & 0.91 & 1.23 & 4.7 &  6.5 & 2935 & $-$2.356 & 0.0273 \\
16:10:30.14$-$23:15:16.8 & 14.367 &    $-$6.3 &    $-$5.7 &   $-$19.4 &   $-$21.3 & 3.1 & 3.2 & 2.9 & 3.1 & --- & 4.6 & --- & 5.1 & 3.04 & 1.87 & 0.89 & 1.23 & 4.6 &  7.0 & 2880 & $-$2.344 & 0.0276 \\
16:13:40.79$-$22:19:46.1 & 14.721 &     --- &     --- &     --- &     --- & 2.2 & 3.3 & 2.3 & 3.5 & --- & --- & --- & --- & 3.40 & 1.92 & 0.89 & 1.37 & 5.6 &  7.5 & 2800 & $-$2.478 & 0.0243 \\
16:14:21.44$-$23:39:14.8 & 14.968 &     --- &     --- &   $-$43.0 &   $-$35.3 & --- & --- & 4.2 & 4.7 & --- & --- & 4.7 & --- & ---  & ---  & ---  & ---  & --- &  7.0 & 2880 & $-$2.584 & 0.0216 \\
16:08:30.49$-$23:35:11.0 & 14.879 &   $-$97.4 &  $-$311.7 &     --- &     --- & 3.4 & 3.7 & --- & --- & --- & 5.7 & --- & --- & ---  & ---  & ---  & ---  & --- &  8.5 & 2550 & $-$2.533 & 0.0229 \\
16:06:48.18$-$22:30:40.1 & 14.926 & $-$1780.0 &  $-$729.8 &     --- &     --- & 4.1 & 5.9 & --- & --- & --- & --- & --- & --- & ---  & ---  & ---  & ---  & --- &  8.5 & 2550 & $-$2.552 & 0.0225 \\
16:07:23.82$-$22:11:02.0 & 15.202 &  $-$671.0 &  $-$431.5 &     --- &     --- & 1.0 & 3.4 & --- & --- & --- & --- & --- & --- & ---  & ---  & ---  & ---  & --- &  7.5 & 2800 & $-$2.670 & 0.0195 \\
16:10:47.13$-$22:39:49.4 & 15.258 &     --- &   $-$55.3 &   $-$28.9 &     --- & 4.8 & 5.1 & 3.3 & 3.5 & --- & --- & --- & --- & ---  & ---  & ---  & ---  & --- &  8.5 & 2550 & $-$2.684 & 0.0191 \\
\hline
16:06:03.75-22:19:30.0 & 15.853 & --- & --- & --- & --- & --- & --- & --- & --- & --- & --- & --- & --- & --- & --- & --- & --- & --- & 12.00 & 1950 & $-$2.886 & 0.0167 \\
16:06:06.29-23:35:13.3 & 16.204 & --- & --- & --- & --- & --- & --- & --- & --- & --- & --- & --- & --- & --- & --- & --- & --- & --- & 10.00 & 2250 & $-$3.031 & 0.0148 \\
16:06:48.18-22:30:40.1 & 14.926 & --- & --- & --- & --- & --- & --- & --- & --- & --- & --- & --- & --- & --- & --- & --- & --- & --- & 8.00 & 2710 & $-$2.552 & 0.0224 \\
16:07:14.79-23:21:01.2 & 16.556 & --- & --- & --- & --- & --- & --- & --- & --- & --- & --- & --- & --- & --- & --- & --- & --- & --- & 10.00 & 2250 & $-$3.172 & 0.0131 \\
16:07:23.82-22:11:02.0 & 15.202 & --- & --- & --- & --- & --- & --- & --- & --- & --- & --- & --- & --- & --- & --- & --- & --- & --- & 11.00 & 2100 & $-$2.630 & 0.0202 \\
16:07:27.82-22:39:04.0 & 16.810 & --- & --- & --- & --- & --- & --- & --- & --- & --- & --- & --- & --- & --- & --- & --- & --- & --- & 11.00 & 2100 & $-$3.273 & 0.0118 \\
16:07:37.99-22:42:47.0 & 16.757 & --- & --- & --- & --- & --- & --- & --- & --- & --- & --- & --- & --- & --- & --- & --- & --- & --- & 10.00 & 2250 & $-$3.252 & 0.0121 \\
16:08:18.43-22:32:25.0 & 16.099 & --- & --- & --- & --- & --- & --- & --- & --- & --- & --- & --- & --- & --- & --- & --- & --- & --- & 10.00 & 2250 & $-$2.989 & 0.0153 \\
16:08:28.47-23:15:10.4 & 15.449 & --- & --- & --- & --- & --- & --- & --- & --- & --- & --- & --- & --- & --- & --- & --- & --- & --- & 11.00 & 2100 & $-$2.729 & 0.0188 \\
16:08:30.49-23:35:11.0 & 14.879 & --- & --- & --- & --- & --- & --- & --- & --- & --- & --- & --- & --- & --- & --- & --- & --- & --- & 9.00 & 2400 & $-$2.521 & 0.0229 \\
16:08:47.44-22:35:47.9 & 15.688 & --- & --- & --- & --- & --- & --- & --- & --- & --- & --- & --- & --- & --- & --- & --- & --- & --- & 9.00 & 2400 & $-$2.844 & 0.0172 \\
16:08:43.44-22:45:16.0 & 18.585 & --- & --- & --- & --- & --- & --- & --- & --- & --- & --- & --- & --- & --- & --- & --- & --- & --- & 11.00 & 2100 & $-$3.983 & 0.0054 \\
16:09:18.69-22:29:23.7 & 18.083 & --- & --- & --- & --- & --- & --- & --- & --- & --- & --- & --- & --- & --- & --- & --- & --- & --- & 11.00 & 2100 & $-$3.782 & 0.0067 \\
16:10:47.13-22:39:49.4 & 15.258 & --- & --- & --- & --- & --- & --- & --- & --- & --- & --- & --- & --- & --- & --- & --- & --- & --- & 9.00 & 2400 & $-$2.672 & 0.0196 \\
16:12:27.64-21:56:40.8 & 17.141 & --- & --- & --- & --- & --- & --- & --- & --- & --- & --- & --- & --- & --- & --- & --- & --- & --- & 10.00 & 2250 & $-$3.406 & 0.0104 \\
16:12:28.95-21:59:36.1 & 16.407 & --- & --- & --- & --- & --- & --- & --- & --- & --- & --- & --- & --- & --- & --- & --- & --- & --- & 11.00 & 2100 & $-$3.112 & 0.0138 \\
16:13:02.32-21:24:28.4 & 17.169 & --- & --- & --- & --- & --- & --- & --- & --- & --- & --- & --- & --- & --- & --- & --- & --- & --- & 10.00 & 2250 & $-$3.417 & 0.0102 \\
16:14:21.44-23:39:14.8 & 14.968 & --- & --- & --- & --- & --- & --- & --- & --- & --- & --- & --- & --- & --- & --- & --- & --- & --- & 7.00 & 2880 & $-$2.584 & 0.0216 \\
16:14:41.68-23:51:05.9 & 16.069 & --- & --- & --- & --- & --- & --- & --- & --- & --- & --- & --- & --- & --- & --- & --- & --- & --- & 11.00 & 2100 & $-$2.977 & 0.0154 \\
\end{longtable}
\end{landscape}
}
%
\end{appendix}

%
%
\begin{appendix}
\section{Classification of stars in the AAOmega field}
\label{USco:appendix_field_stars}

The full sample of over 600 stars in the AAOmega field assigned to a fibre 
in addition to the USco member candidates is available electronically
(Table \ref{tab_USco:field_stars}).
An example of the table is provided below for guidance only.
Columns 1 and 2 give the coordinates (J2000); Columns 3--7
the $ZYJHK$ photometry, Columns 8--9 the proper motions (in mas/yr)
in right ascension and declination, respectively; Column 10 the
spectral types based on the visual comparison with the templates
listed below; and Column 11 the name give prior to the fibre
assignment.

\begin{figure}
   \includegraphics[angle=0, width=\linewidth]{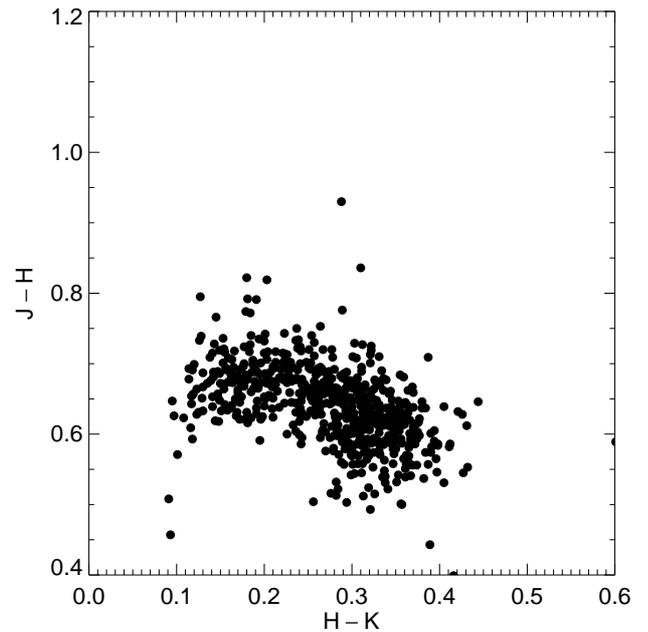}
   \caption{($H-K$,$J-H$) two-colour diagram for stars in the AAOmega field.}
   \label{fig_USco:field_stars_CCD}
\end{figure}

\begin{figure*}
   \includegraphics[angle=0, width=0.49\linewidth]{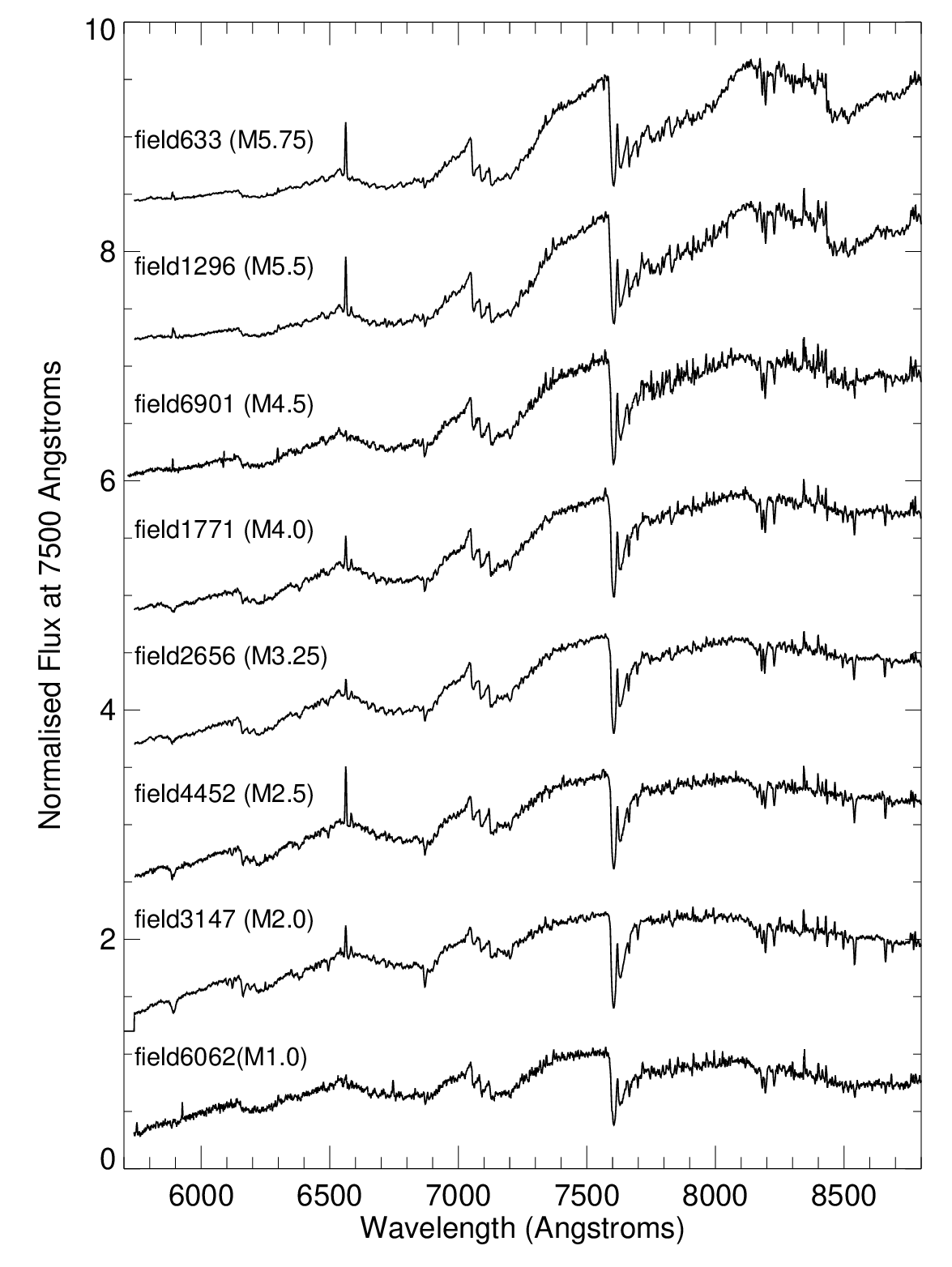}
   \includegraphics[angle=0, width=0.49\linewidth]{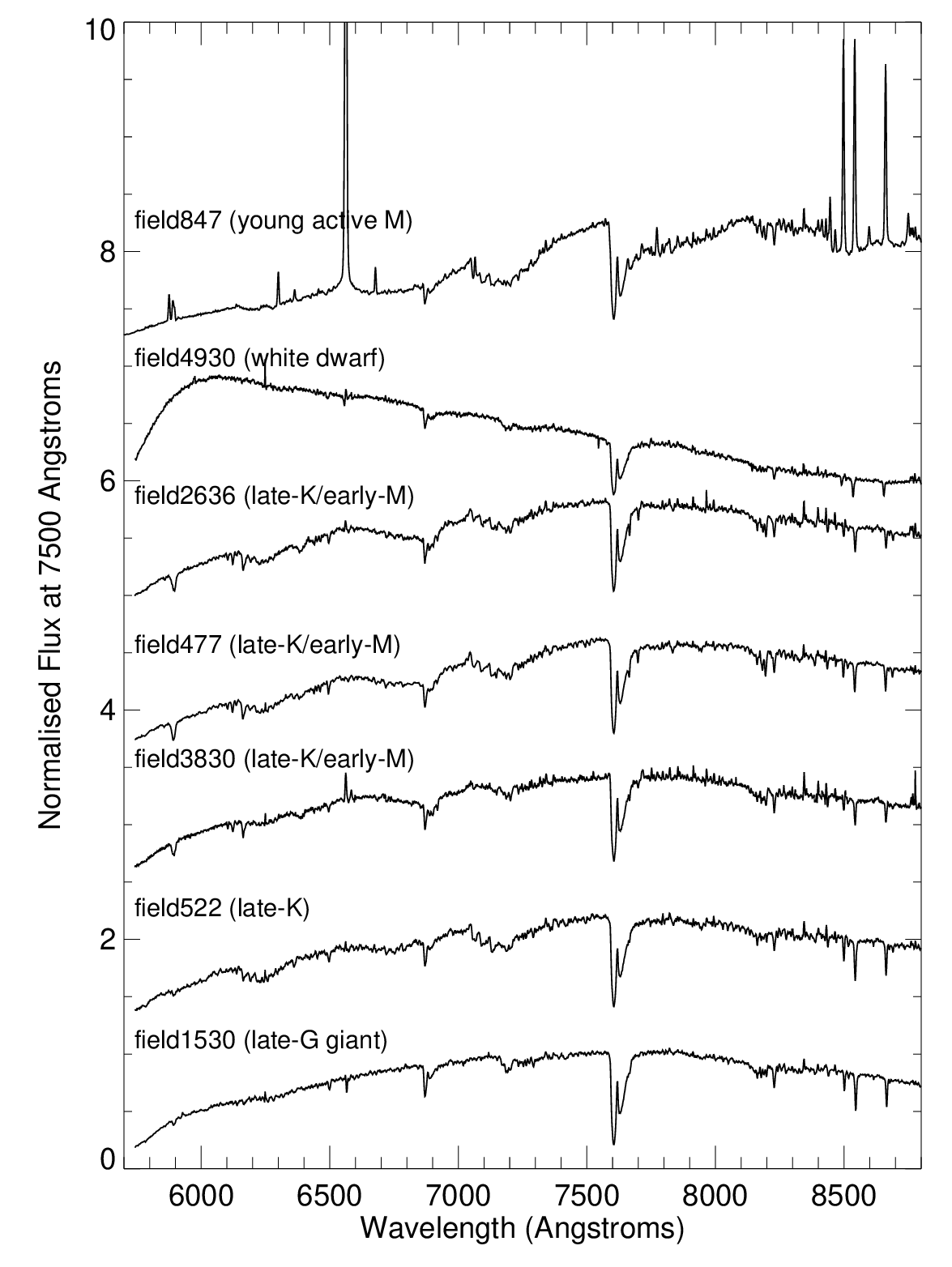}
   \caption{Example of optical spectra obtained with AAT/AAOmega for
for stars in the AAOmega field not selected as photometric candidates. 
Spectra are normalised at 7500\AA{} and offset for clarity.
{\it{Left:}} A sample of field M dwarfs with spectral types ranging
from M1 to M6\@.
{\it{Right:}} A sample of late-K/early-M dwarfs, giants, white dwarfs, 
and young active M dwarfs.
}
   \label{fig_USco:spec_others}
\end{figure*}

This sample can be divided into several groups, each of them
characterised by one template with a good quality spectrum
(see Fig.\ \ref{fig_USco:spec_others}). Spectral types of
M dwarfs should be accurate to half a subclass whereas others 
are tentative spectral types. The templates are:

\begin{itemize}
\item[$\bullet$] UGCS J160709.43$-$234031.8 or field6062 classified as M1.5
\item[$\bullet$] UGCS J160709.43$-$234031.8 or field3147 classified as M2
\item[$\bullet$] UGCS J160709.43$-$234031.8 or field4452 classified as M2.5
\item[$\bullet$] UGCS J160709.43$-$234031.8 or field2656 classified as M3--M3.5
\item[$\bullet$] UGCS J160709.43$-$234031.8 or field1771 classified as M4
\item[$\bullet$] UGCS J160709.43$-$234031.8 or field6901 classified as M4.5
\item[$\bullet$] UGCS J160709.43$-$234031.8 or field1296 classified as M5.25--5.5
\item[$\bullet$] UGCS J160709.43$-$234031.8 or field633 classified as M5.75--M6
\item[$\bullet$] Four objects classified as dM6$+$ i.e.\ later than M6
\item[$\bullet$] UGCS J160634.44$-$231517.7 or field3830 is classified as 
a late-K/early-M
\item[$\bullet$] UGCS J160709.43$-$234031.8 or field2636 is classified as 
a late-K/early-M
\item[$\bullet$] UGCS J160733.42$-$224609.4 or field477 is classified as
late-K/early-M and show a different level of reddening than field3830\@.
The position of these objects in the ($H-K$,$J-H$) colour-colour diagram
(Fig.\ \ref{fig_USco:field_stars_CCD}) corroborate the classification as 
dwarfs rather than giants \citep[e.g. Fig.\ 5 in][]{dobbie02c}. However,
we can not rule our that some giants are  included in this sample
\item[$\bullet$] UGCS J160544.57$-$231310.6 or field522 is classified as 
a late-K dwarf or giant
\item[$\bullet$] UGCS J160632.31$-$224817.2 or field1530 is classified as 
a late-G giant. Sources in this group show similar features but may have 
different levels of reddening
\item[$\bullet$] Two white dwarfs: UGCS J160757.34$-$231248.5 (field4930)
and UGCS J160917.16$-$231020.1 (field1861)
\item[$\bullet$] Three young active flare M dwarfs, see Sect.\ \ref{USco_AAOmega:Field_accreting}
\item[$\bullet$] 18 objects unclassified
\item[$\bullet$] 11 objects with no signal
\end{itemize}

%
%
%
%

%
%
%
%

%
%
%
\begin{table*}
\caption{Coordinates (J2000), $ZYJHK$ photometry, and proper motions
(in arcsec/yr) of stars in the AAOmega field ordered by increasing 
$Z$ magnitude.}
\label{tab_USco:field_stars}
\centering
\begin{tabular}{c c c c c c c c c c c}
\hline
\hline
RA & dec & $Z$ & $Y$ & $J$ & $H$ & $K$ & $\mu_\alpha \cos\delta$ & $\mu\delta$ & SpT  & Name \\
h s m & $^\circ$ ' '' & mag & mag & mag & mag & mag & mas/yr & mas/yr &  &  \\
\hline
16:07:03.10 &  $-$23:31:46.3 & 12.024 & 11.539 & 10.982 & 10.393 &  9.792 &    10.21 &   $-$24.90 & M3--M3.5    & field229     \cr
\ldots{}    &  \ldots{}      & \ldots{} & \ldots{} & \ldots{} & \ldots{} & \ldots{} & \ldots{} & \ldots{} & \ldots{} & \ldots{} \cr
\ldots{}    &  \ldots{}      & \ldots{} & \ldots{} & \ldots{} & \ldots{} & \ldots{} & \ldots{} & \ldots{} & \ldots{} & \ldots{} \cr
16:15:28.01 &  $-$23:08:09.6 & 17.489 & 16.816 & 16.263 & 15.698 & 15.339 &   $-$67.67 &    57.84 & M5.25--M5    & field31105   \cr
\hline
\end{tabular}
\end{table*}
\end{appendix}

%
%
\begin{appendix}
\section{Proper motion non members}
%
%
\begin{table*}
\caption{Proper motion non members from \citet{lodieu07a}
observed during our AAOmega spectroscopic follow-up.}
\label{tab_USco:PM_NM_L07a}
\centering
\begin{tabular}{l c c c c c c c c c c}
\hline
\hline
USco\,J\ldots{}      & $Z$   & $\mu_\alpha \cos\delta$ & $\mu_{\delta}$ & SpT   & H$\alpha$ EW & Na {\small{I}} EW & Li & PM & Memb & UCAC PM \\
                     & mag   & mas/yr                  & mas/yr         &       & \AA{}        &    \AA{}          &    &    &      & mas/yr  \\
\hline
160827.38$-$221729.4 & 12.76 &  $+$0.38  &  $-$8.08  & M5.0  & $-$9.2   &  1.8 2.0 (3.8) & Y  & $\geq$2$\sigma$   &  Memb  & $-$10.6 $-$19.4 \\
160834.60$-$221156.0 & 13.77 &  $-$7.13  &  $-$5.24  & M4.5  & $-$12.2  &  1.6 1.9 (3.5) & Y? & $\geq$2$\sigma$   &  Memb  &  --- \\
160907.80$-$233954.6 & 13.44 &  $-$17.08 &  $-$3.69  & M5.5  & $-$11.7  &  1.5 2.0 (3.5) & Y? & $\geq$2$\sigma$   &  Memb  &  --- \\
160908.88$-$221747.0 & 14.12 &  $-$36.50 &  $-$29.63 & M5.75 & $-$17.4  &  2.2 2.6 (4.8) & Y? & $\geq$2.5$\sigma$ &  Memb  &  --- \\
161000.23$-$231219.4 & 15.95 &  $-$33.51 &  $-$14.21 & M6.0  & $-$8.5   &  2.8 3.4 (6.2) & Y  & $\geq$2$\sigma$   &  NM    &  --- \\
161019.48$-$233109.0 & 14.84 &  $-$5.10  &  $-$2.45  & M5.75 & $-$10.0  &  1.3 1.8 (3.1) & Y  & $\geq$2$\sigma$   &  Memb  &  --- \\
161412.41$-$221913.3 & 12.47 &  $+$8.80  &  $+$0.24  & M4.5  & $-$7.0   &  1.6 1.7 (3.3) & Y  & $\geq$3$\sigma$   &  Memb  &  $-$9.1 $-$23.8 \\
161516.07$-$234510.5 & 13.59 &  $-$11.05 &  $+$3.86  & M5.0  & $-$13.5  &  1.6 1.9 (3.5) & Y  & $\geq$2.5$\sigma$ &  Memb  &  --- \\
161538.48$-$234156.1 & 13.40 &  $-$32.06 &  $-$16.17 & M5.0  & $-$8.4   &  1.8 2.1 (3.9) & Y  & $\geq$2$\sigma$   &  Memb  &  --- \\
161620.16$-$234414.4 & 12.08 &  $-$33.21 &  $-$11.33 & M5.0  & $-$9.8   &  1.8 2.2 (4.0) & Y  & $\geq$2.5$\sigma$ &  Memb  & $-$12.2 $-$27.4 \\
\hline
\end{tabular}
\end{table*}
\end{appendix}

%
%
%
\begin{appendix}
\section{Candidates with proper motion and equivalent widths satisfying membership}
%
%
\begin{table*}
\caption{List of 50 candidates with proper motion within 3$\sigma$ from
the cluster mean motion and with equivalent widths measurements
of H$\alpha$ and the Na {\small{I}} doublet consistent with membership.
The eight proper motion non members reclassified as candidate members 
and listed in Table \ref{tab_USco:PM_NM_L07a} are not included.
We have added at the top of the table the two accreting sources 
with proper motions consistent with USco for consistency.
}
\label{tab_USco:missed_Memb}
\centering
\begin{tabular}{l c c c c c c c c c c c}
\hline
\hline
R.A.\       &  dec     &  $Z$   &   $Y$  &   $J$  &  $H$   &  $K$   &  $\mu_\alpha \cos\delta$ & $\mu_{\delta}$ & H$\alpha$ EW & Na {\small{I}} EW & SpT  \\
h m s  &  $^\circ$ ' '' &  mag  &   mag  &   mag  &  mag   &  mag   &  mas/yr                  &      mas/yr    &   \AA{}      &    \AA{}          &      \\
\hline
16:07:29.59 & $-$23:08:22.4 & 12.815 & 12.365 & 11.749 & 10.970 & 10.188 &   $-$30.12 &   $-$29.77 & $-$158 & 5.1 & M3.0 \\
16:14:50.31 & $-$23:32:40.0 & 12.758 & 12.218 & 11.605 & 10.803 & 10.067 &   $-$18.21 &   $-$16.37 & $-$108 & 3.0 & M4.5 \\
\hline
16:05:54.13 & $-$22:57:20.1 & 13.720 & 13.392 & 13.008 & 12.337 & 12.124 &   $-$17.98 &     2.37 &   $-$2.9 & 3.3 & K7.0 \\
16:05:59.16 & $-$23:13:30.3 & 14.796 & 14.376 & 13.891 & 13.242 & 12.980 &   $-$20.93 &   $-$19.70 &   $-$7.4 & 4.2 & M4.0 \\
16:06:21.43 & $-$23:06:39.7 & 14.434 & 14.052 & 13.624 & 12.952 & 12.708 &    $-$2.58 &   $-$22.89 &   $-$2.5 & 3.7 & M2.0 \\
16:06:33.11 & $-$22:39:07.6 & 14.318 & 14.015 & 13.579 & 12.942 & 12.701 &     5.08 &    $-$7.73 &   $-$8.2 & 3.5 & M2.0 \\
16:06:40.45 & $-$22:24:55.0 & 14.935 & 14.467 & 13.972 & 13.307 & 13.005 &    $-$0.63 &    $-$1.36 &  $-$14.5 & 4.9 & M4.0 \\
16:06:42.49 & $-$22:47:14.7 & 15.664 & 15.159 & 14.666 & 14.065 & 13.759 &     4.42 &    $-$6.36 &  $-$22.8 & 5.0 & M4.0 \\
16:06:47.21 & $-$22:24:38.6 & 15.325 & 14.829 & 14.292 & 13.672 & 13.358 &   $-$14.01 &    $-$2.72 &  $-$18.9 & 4.5 & M3.2 \\
16:06:55.26 & $-$22:47:09.0 & 16.083 & 15.555 & 15.053 & 14.436 & 14.110 &    $-$8.30 &    $-$0.35 &  $-$34.4 & 4.6 & M5.5 \\
16:07:03.05 & $-$23:31:46.2 & 12.024 & 11.539 & 10.982 & 10.393 &  9.792 &    10.21 &   $-$24.90 &  $-$12.5 & 3.5 & M3.2 \\
16:07:06.86 & $-$22:52:21.9 & 15.233 & 14.776 & 14.307 & 13.634 & 13.367 &   $-$34.77 &   $-$42.90 &   $-$6.3 & 4.2 & M4.0 \\
16:07:11.28 & $-$23:01:10.8 & 15.803 & 15.298 & 14.821 & 14.221 & 13.924 &     2.35 &   $-$14.59 &  $-$19.9 & 4.6 & M4.5 \\
16:07:23.23 & $-$23:42:24.5 & 15.055 & 14.620 & 14.156 & 13.513 & 13.238 &   $-$26.84 &   $-$26.08 &   $-$2.1 & 3.1 & M3.2 \\
16:07:24.27 & $-$23:02:27.4 & 15.611 & 15.080 & 14.574 & 14.017 & 13.726 &   $-$16.65 &   $-$16.85 &  $-$19.7 & 4.5 & M4.5 \\
16:07:39.59 & $-$23:12:13.6 & 14.472 & 14.101 & 13.632 & 12.953 & 12.716 &     1.26 &   $-$18.52 &   $-$2.0 & 3.5 & M2.0 \\
16:07:41.59 & $-$23:39:09.7 & 15.503 & 15.008 & 14.542 & 13.939 & 13.673 &     7.17 &   $-$17.27 &   $-$2.2 & 3.8 & M4.0 \\
16:07:46.74 & $-$22:25:04.6 & 15.416 & 14.955 & 14.467 & 13.852 & 13.584 &     6.87 &   $-$19.85 &  $-$13.5 & 4.7 & M4.0 \\
16:07:48.11 & $-$22:45:29.8 & 14.196 & 13.817 & 13.403 & 12.722 & 12.498 &    $-$1.77 &   $-$20.76 &   $-$1.4 & 3.2 & M0.0 \\
16:07:49.13 & $-$22:30:42.3 & 15.166 & 14.738 & 14.240 & 13.632 & 13.338 &   $-$12.26 &   $-$10.90 &  $-$11.8 & 4.0 & M4.0 \\
16:07:53.62 & $-$23:41:27.4 & 15.528 & 15.008 & 14.535 & 13.854 & 13.565 &   $-$10.06 &    $-$1.51 &   $-$1.3 & 4.9 & M4.0 \\
16:07:57.23 & $-$22:31:06.0 & 15.178 & 14.743 & 14.261 & 13.613 & 13.313 &   $-$25.67 &   $-$18.50 &  $-$10.0 & 4.1 & M4.0 \\
16:08:09.60 & $-$22:27:25.8 & 14.549 & 14.180 & 13.749 & 13.088 & 12.873 &    $-$2.21 &    $-$1.00 &   $-$3.6 & 2.9 & M0.0 \\
16:08:10.35 & $-$22:54:55.4 & 15.397 & 14.909 & 14.446 & 13.786 & 13.503 &    $-$0.28 &     2.79 &  $-$10.8 & 4.0 & M3.2 \\
16:08:11.54 & $-$22:34:52.5 & 15.880 & 15.406 & 14.894 & 14.337 & 14.036 &    $-$4.58 &    $-$7.65 &  $-$20.6 & 4.3 & M4.5 \\
16:08:23.42 & $-$22:35:04.5 & 15.083 & 14.683 & 14.191 & 13.527 & 13.282 &     4.18 &    $-$7.07 &   $-$4.9 & 3.5 & M2.5 \\
16:08:26.24 & $-$22:49:51.9 & 14.963 & 14.514 & 14.051 & 13.350 & 13.097 &    $-$1.15 &    $-$6.07 &   $-$3.7 & 4.0 & M2.5 \\
16:08:40.97 & $-$22:46:32.6 & 15.307 & 14.785 & 14.282 & 13.630 & 13.343 &     5.34 &    $-$8.09 &   $-$9.7 & 4.8 & M4.0 \\
16:08:41.92 & $-$23:08:02.7 & 13.128 & 12.748 & 12.314 & 11.614 & 11.373 &   $-$30.05 &   $-$31.33 &    0.0 & 4.4 & M2.0 \\
16:08:47.94 & $-$23:24:19.3 & 15.905 & 15.414 & 14.858 & 14.294 & 13.977 &    $-$4.41 &   $-$34.91 &   $-$1.8 & 4.7 & M4.0 \\
16:08:52.43 & $-$22:17:59.0 & 13.817 & 13.479 & 13.068 & 12.432 & 12.223 &    $-$3.45 &    $-$0.39 &   $-$1.0 & 3.8 & M0.0 \\
16:08:53.34 & $-$23:07:41.5 & 15.516 & 15.046 & 14.506 & 13.900 & 13.580 &    $-$1.96 &   $-$18.55 &   $-$5.0 & 4.6 & M4.0 \\
16:08:53.81 & $-$22:35:01.5 & 14.008 & 13.620 & 13.166 & 12.522 & 12.299 &     5.34 &   $-$25.03 &   $-$1.7 & 3.7 & M3.2 \\
16:08:58.04 & $-$22:43:04.2 & 15.478 & 14.942 & 14.463 & 13.816 & 13.506 &     3.09 &    $-$5.86 &  $-$14.1 & 4.3 & M4.5 \\
16:08:58.07 & $-$22:17:52.2 & 15.475 & 14.971 & 14.482 & 13.893 & 13.545 &    $-$6.81 &   $-$19.49 &   $-$8.5 & 4.1 & M4.0 \\
16:09:03.40 & $-$22:16:45.4 & 14.418 & 14.062 & 13.646 & 12.971 & 12.762 &   $-$28.06 &   $-$21.50 &   $-$3.6 & 3.5 & M0.0 \\
16:09:07.21 & $-$22:21:15.8 & 14.257 & 13.837 & 13.359 & 12.670 & 12.409 &   $-$33.04 &   $-$19.10 &   $-$2.1 & 4.7 & M3.2 \\
16:09:07.81 & $-$22:21:05.7 & 16.974 & 16.383 & 15.840 & 15.227 & 14.912 &     3.01 &   $-$24.30 &  $-$31.4 & 4.0 & M4.5 \\
16:09:12.01 & $-$23:00:57.3 & 16.119 & 15.578 & 15.100 & 14.482 & 14.156 &   $-$27.23 &   $-$14.46 &   $-$7.5 & 3.9 & M3.2 \\
16:09:15.41 & $-$22:59:01.8 & 16.090 & 15.528 & 15.034 & 14.424 & 14.079 &   $-$24.41 &   $-$22.02 &   $-$8.4 & 4.7 & M4.0 \\
16:09:16.66 & $-$23:54:48.6 & 14.891 & 14.330 & 13.824 & 13.145 & 12.868 &   $-$18.43 &     2.49 &   $-$8.9 & 4.6 & M4.5 \\
16:09:47.50 & $-$22:15:41.0 & 15.888 & 15.392 & 14.886 & 14.267 & 13.985 &    $-$0.38 &   $-$44.42 &   $-$5.2 & 4.7 & M4.5 \\
16:10:03.71 & $-$22:41:34.6 & 16.113 & 15.514 & 15.014 & 14.428 & 14.094 &    $-$4.66 &   $-$49.18 &   $-$4.9 & 5.0 & M4.0 \\
16:10:10.76 & $-$23:08:27.3 & 14.520 & 14.132 & 13.661 & 12.937 & 12.696 &   $-$14.65 &     3.86 &   $-$1.8 & 3.8 & M2.0 \\
16:10:13.43 & $-$22:43:41.7 & 15.763 & 15.197 & 14.690 & 14.103 & 13.746 &     0.00 &     0.00 &   $-$2.1 & 4.4 & M4.5 \\
16:10:55.38 & $-$23:19:53.5 & 15.374 & 14.902 & 14.405 & 13.752 & 13.474 &    $-$4.53 &    $-$0.06 &   $-$6.7 & 4.4 & M4.0 \\
16:12:57.88 & $-$23:58:26.5 & 16.893 & 16.210 & 15.696 & 15.019 & 14.680 &   $-$17.90 &   $-$26.02 &   $-$4.3 & 4.6 & M4.5 \\
16:14:02.69 & $-$22:57:26.1 & 14.820 & 14.300 & 13.815 & 13.176 & 12.876 &   $-$11.13 &   $-$10.24 &   $-$2.0 & 4.8 & M3.2 \\
16:15:12.62 & $-$23:56:14.6 & 14.299 & 13.813 & 13.346 & 12.693 & 12.378 &     3.70 &   $-$10.74 &   $-$4.8 & 3.9 & M3.2 \\
16:16:03.50 & $-$23:41:03.6 & 15.156 & 14.572 & 14.063 & 13.441 & 13.147 &    $-$2.97 &   $-$10.99 &   $-$6.4 & 4.9 & M4.5 \\
16:16:52.66 & $-$22:29:48.8 & 16.014 & 15.431 & 14.853 & 14.268 & 13.932 &   $-$19.27 &   $-$10.57 &   $-$8.6 & 4.6 & M4.5 \\
16:16:54.25 & $-$23:56:53.4 & 16.712 & 16.081 & 15.558 & 14.879 & 14.543 &   $-$10.83 &    $-$5.65 &   $-$3.2 & 4.8 & M4.5 \\
\hline
\end{tabular}
\end{table*}
\end{appendix}

\end{document}